\documentclass[letter,12pt]{revtex4-2}

\usepackage{bm}

\usepackage{hyperref}
\usepackage{siunitx}
\usepackage{mathtools}
\usepackage{bbold}
\usepackage[normalem]{ulem}
\usepackage{enumerate}
\usepackage[dvipsnames]{xcolor}
\def\be{\begin{equation}}
\def\ee{\end{equation}}
\def\e#1{\label{#1}\end{equation}}
\def\bea{\begin{eqnarray}}
\def\eea{\end{eqnarray}}
\def\ea#1{\label{#1}\end{eqnarray}}

\def\bem#1{\begin{mathletters}\label{#1}}
\def\eml{\end{mathletters}}

\newcommand{\ket}[1]{\left| #1 \right>}

\def\braket#1#2{{\langle#1|#2\rangle}}

\def\4#1{{\boldsymbol{#1}}}
\def\8#1{{\widetilde{#1}}}
\def\bse{\begin{subequations}}
\def\ese{\end{subequations}}

\begin{document}

\title{The case against entanglement improved measurement precision}

\author{Liam P. McGuinness}
\affiliation{Laser Physics Centre, Research School of Physics, Australian National University, Acton, Australian Capital Territory 2601, Australia \\ \textnormal{Email: \href{mailto:liam@grtoet.com}{liam@grtoet.com}} }

\begin{abstract}
It is widely accepted that quantum entanglement between otherwise independent sensors can yield a measurement precision beyond that achievable when the same resources are employed without entanglement \cite{Helstrom1969, Holevo1973a, Caves1980a, Caves1981, Wootters1981, Yurke1986, Wu1986, Xiao1987, Slusher1987, Shapiro1989, Wineland1992, Polzik1992, Kitagawa1993, Braunstein1994, Wineland1994, Sanders1995, Bollinger1996, Ou1997, Dowling1998,Soerensen1998, Brif1999, Childs2000, Fleischhauer2000, Meyer2001, Geremia2003, Giovannetti2004, Kok2004, Leibfried2004, Leibfried2005, Giovannetti2006, Nagata2007, Appel2009, Gross2010, Leroux2010, Zwierz2010,DemkowiczDobrzanski2012, Zwierz2012,Aasi2013,Pezze2018,Tse2019,Casacio2021}. Here we show that theoretical proofs of entanglement enhanced metrology are based on a misinterpretation of \emph{can't} theorems as \emph{can} theorems. In concert, we dissect claims of an experimental precision beyond the classical limits and detail where comparisons are misleading, incomplete or incorrect to show that the precision of optimised measurements which forgo entanglement has not been surpassed. In doing so, we highlight a significant discrepancy between experimentally reported uncertainties and the current predictions of quantum measurement theory. The discrepancy can be resolved by introducing a simple physical principle which demonstrates better agreement to empirical evidence. We thus provide viable avenues as to where standard interpretations of quantum mechanics should be modified in order to better predict measurement outcomes.

\end{abstract}

\maketitle

\pagestyle{plain}
The central thesis of this paper is that with access to $N$ identical particles (atoms or photons) and $T$ measurement time, the uncertainty of any measuring device cannot surpass $1/\sqrt{N}$ times the optimal uncertainty obtained with a single sensor in time $T$, even allowing for entanglement. 
For concreteness, we use the example of Ramsey interferometry with $S=1/2$ spins \cite{Ramsey1950}, whilst emphasizing that the results are not restricted to this example and can be extended to quantum measurements in general through mathematical equivalencies. Indeed the first proposals for entanglement enhanced metrology (EEM) with more than $1/\sqrt{N}$ improvement were based on Mach-Zehnder interferometry with photons \cite{Yurke1986, Wineland1994} which is regarded as equivalent to Ramsey interferometry with spins \cite{Caves1981, Shapiro1989}. Specifically, we assert that for certain classes of unknown parameters $x$, the best estimate of $x$ (denoted $\hat{x}$) achievable in time $T$ with a Ramsey interferometer, has an uncertainty bounded by:
\begin{equation}
\Delta\hat{x}_{\mathrm{SQL}} > \frac{\lambda}{\sqrt{N}T}, \quad \quad \mathrm{(SQL)}
\label{eq:SQL}
\end{equation}
sometimes called the standard quantum limit (SQL), where $N$ is the number of spins in the Ramsey interferometer and $\lambda$ is a constant factor determined by physical constants such that $\Delta\hat{x}$ has the same units as $x$ (see \ref{App:SQL} for details and derivation). This bound can be compared to the Heisenberg limit (HL) in quantum metrology \cite{Bollinger1996, Childs2000, Aharonov2002, Zwierz2010}:
\begin{equation}
\Delta\hat{x}_{\mathrm{HL}} > \frac{\lambda}{N T},	\quad \quad \mathrm{(HL)}
\label{eq:HL}
\end{equation}
which is assumed approachable when the spins are entangled and we have taken the liberty to drop any possibility of equality. As $N \geq 1$, the lower bound of the HL is never worse than the bound of the SQL, coinciding only for $N = 1$, where no entanglement is possible. Comparison of these bounds has led to claims that with sufficient experimental effort, entanglement can surpass the precision attainable without entanglement. In contrast, our thesis posits that Eq.\,\eqref{eq:SQL} \emph{fundamentally} limits the achievable uncertainty, which can never be surpassed, irrespective of overcoming technical limitations. Furthermore Eq.\,\eqref{eq:SQL} is a loose bound and can be tightened (see \ref{App:1}).

Now we have outlined our thesis, the next sections provide supporting arguments. We take three approaches:
\begin{enumerate}[i)]
\item Discussion of logical gaps in the derivation of the Heisenberg limit as a tight bound on the minimum uncertainty. We explain that saturability of the HL has not been proven, and all rigorous mathematical statements on Heisenberg limited estimation contain assumptions. This is an important clarification, since a proof that the HL can be saturated would invalidate the premise of this paper.
\item A literature search for any experimental (dis)confirmation of the SQL. We review experimentally demonstrated claims of an uncertainty below the SQL. Again, this paper's thesis would be voided by the existence of any such evidence.
\item The introduction of new analysis and a basic principle which implies the central thesis. The absence of any theoretical or experimental verification of Heisenberg limited metrology does not in itself support the central thesis. Here we aim to shift the argumentative balance by introducing theoretical principles which provide testable experimental predictions.\\
\end{enumerate}

\renewcommand{\thesubsection}{\arabic{subsection}}

\subsection{Derivation and interpretation of the Heisenberg limit}\label{Sec:Assumptions}
First we address the belief that quantum entanglement can provide a measurement precision beyond the classical limits\footnote{Throughout this paper ‘classical’ means ‘non-entangled’, although to be clear the classical limits are quantum mechanical in that they are derived from a quantum mechanical description of independent spins and uncertainty relations. We compare non-entangled states against entangled states since a non-entangled state has a precise mathematical definition - it can be written as a separate product of independent systems, which avoids arguments about what constitutes `classical' or `quantum', and makes the arguments rigorous.}, since as long as reader is convinced that this is a foregone conclusion, subsequent arguments will fall on deaf eyes. 
The first surprising observation is that there is no intrinsic conflict between the Heisenberg limit and the standard quantum limit, they can both be true since Eq.\,\eqref{eq:SQL} implies Eq.\,\eqref{eq:HL}, although not the converse. In other words, our thesis holds that the Heisenberg limit is indeed correct, but that additional tighter bounds also apply\footnote{By way of explanation, here is an analogy, which while not perfect, is illustrative of the main issues at stake and arguments in play. \emph{You manage a car racing team with a well-defined goal: to develop a car that can perform one circuit of a specific racing track in the shortest possible time. Through years of testing and improvements your team has finally produced a car that you are satisfied with... until an engineer comes to you with some modifications. ``Great, let's hear them!'' you say. The engineer produces a table with two columns. On the left, is the top speed of the current car and on the right is the calculated top speed of the modified car - significantly faster. ``Not only that',' he adds, ``we have tested the modifications and recorded a top speed close to our calculations. This proves we can achieve a lap-time faster than possible with the current car.''} Similar arguments are currently used in the quantum metrology community. The HL is an uncertainty bound obtained from the `top speed' of evolution during one part of a quantum measurement, which is provably greater than the `top speed' obtained without entanglement. As the manager, would you like to know the lap-time that the modified car actually achieves? And, under what assumptions is the engineers `proof' valid?}. 

Despite this clarification, the suggestion that the SQL is a fundamental uncertainty limit is met with deep resistance in the quantum metrology community. Any quantum physicist worth their salt will tell you that it can be surpassed, and most argue that the Heisenberg limit can be saturated. So where does Eq.\,\eqref{eq:HL} and the associated belief come from?

It is a multi-part answer.

\begin{enumerate}
\item	By entangling an $N$-spin ensemble, an $N$-fold enhancement in the quantum phase evolution of the ensemble, as compared to the phase evolution of a single spin, is obtained \cite{Bollinger1996, Giovannetti2004, Kok2004}. This has been experimentally verified \cite{Leibfried2004, LouchetChauvet2010, Gross2010, Unden2016} provided basic assumptions on converting quantum mechanical phase into an observable.
\item 	From readout of the quantum phase $\phi$, the statistical (but not systematic) error in $\Delta \hat{x}$ must be lower bounded by Eq.\,\eqref{eq:HL} \cite{Holevo1973a}. This is known as the quantum Cram\'er-Rao lower bound \cite{Helstrom1969}, which is related to quantum Fisher information and entropy \cite{Braunstein1994, Wootters1981} and for some reason the equality in the ‘greater than or equal to’ persists, even though it is clear that equality can never be attained.
\item  It is generally held that the measurement uncertainty can be made arbitrarily close to the HL, dependent on technical improvements in spin initialisation, entanglement fidelities, decoherence rates, readout techniques, and cancellation of systematic errors to which there appears no fundamental limitation. Alternatively, if the entanglement overheads are fixed and decoherence overcome, then for a long enough interaction time, it is thought that the uncertainty will surpass the standard quantum limit and asymptote to the Heisenberg limit.
\end{enumerate}
The following points are not required, but as they seem to have influenced people's beliefs they are included.
\begin{enumerate}
\setcounter{enumi}{3}
\item 	Experimentally, a measurement uncertainty of $\Delta \hat{x} = \alpha \lambda /N$ has been demonstrated for $1 < \alpha < \sqrt{N}$ \cite{Leibfried2004, Gross2010, Shaniv2018}. By excluding the measurement time from the analysis, one can obtain an uncertainty beyond the classical limit solely in terms of the number of measurements per spin (or number of spins if, for some arbitrary reason, we assume only one preparation and measurement per spin is allowed). Note, since we are already disregarding units, Eq.\,\ref{eq:SQL} also allows such an uncertainty without entanglement when $T>1$.
\item 	Experimentally, a measurement uncertainty of $\Delta \hat{x} = \alpha \lambda /(N\sqrt{T})$ has been demonstrated for $1 < \alpha < \sqrt{N}$ \cite{Xiao1987, Leroux2010}. By requiring that multiple identical measurements are performed on each spin and thereby restricting the improvement to $1/\sqrt{T}$ an uncertainty improvement is achieved with entanglement.
\item	Uncertainties that scale better than Eq.\,\eqref{eq:SQL} have been proposed \cite{Luis2004, Beltran2005, Boixo2007, Roy2008, Borregaard2013, Rosenband2013} and demonstrated \cite{Napolitano2011, Santagati2019, Schmitt2021}. If the duration of this scaling is indefinite, then at some point the uncertainty must beat Eq.\,\eqref{eq:SQL}.
\end{enumerate}

Points 1 and 2 are rigorous mathematical statements that follow from the Schr\"odinger equation and the postulates of quantum mechanics concerning measurement observables. Point 3 on the other-hand, which assumes saturation of the Heisenberg limit, contains implicit assumptions which have not been rigorously derived or tested (an early `\emph{gedanken} experiment' proof contains similar assumptions \cite{Caves1980a}). Actually, that's it. Entanglement advantages in quantum metrology arise by considering one part of a quantum measurement in isolation -- phase accumulation -- bounding the uncertainty from that analysis and adding some assumptions that the bound can be achieved. More pointedly, this assumes that each of the technical improvements listed in Point 3 are independent of each other and can be simultaneously improved without any fundamental trade-off's; for example, that entanglement does not introduce any systematic errors. This has led to the promulgation of a central claim in quantum metrology -- that entanglement provides a fundamental sensing advantage -- which has been reinforced by theoretical analyses and experiments that give the impression of surpassing the SQL, without in fact doing so (Points 4 -- 6).

We emphasise that the claim of EEM is based on a theoretical bound, which cannot be achieved. Eq.\,\ref{eq:HL} is a \emph{can't} equation, not a \emph{can} equation. In treating the bound for all practical purposes as a strict equality, the quantum metrology community has committed a huge fallacy -- forgoing the opportunity to investigate the existence of stricter bounds. And this indeed is a question worthy of investigation. We conclude this section by noting that until now we have only highlighted the assumptions that formalise an improvement beyond the standard quantum limit, but we have not provided any refutation of these assumptions. On it's own this is a weak first foray, since those assumptions seem reasonable. The reader might further add: \textit{it would be bloody surprising if some unknown physical law prevented the Heisenberg limit from being experimentally approached. Furthermore the assumptions have already been validated since dozens of papers have experimentally shown an improvement beyond the classical limit}.\\


The next two sections aim to shift that perception dramatically.

\subsection{Experimental tests of the Heisenberg limit}\label{Sec:Experiment}
The central thesis yields predictions which can be experimentally disconfirmed. In particular, Eq.\,\eqref{eq:SQL} provides a testable no-go theorem on the uncertainty that $N$ resources can provide in a given amount of time. So why don’t we check the experiments to see what uncertainty per unit time can be achieved? In this section we review experimentally reported uncertainties and contend that not only is a thorough analysis needed, but moreso broad dissemination of the findings. The reason being that most people's understanding of the current state-of-the-art is severely out of step with reality. For example, many opponents of the central thesis argue that it has been experimentally disproven; often this assertion being proffered by the very authors of said works. If such beliefs are mistaken, then effort should be taken to correct those misconceptions, especially since they are so widespread. While we focus on atomic clock literature in this section, a more detailed review on the measurement precision achieved in other contexts is provided in \ref{App:Experiment}, where we outline several loopholes which are commonly used to give an impression of exceeding the classical limits without in fact achieving such a result.

At risk of spoiling the surprise, our review finds no experimental demonstration of an uncertainty beyond Eq.\,\eqref{eq:SQL}. This lack of \emph{any} experimental evidence is particularly striking when we note that the standard quantum and Heisenberg limits have been experimentally investigated for over 30 years. In the field of atomic clocks particularly, there are strong incentives to surpass the standard quantum limit and these incentives have been reflected in experimental effort. Notwithstanding their immense practical import, atomic time standards are salient in that the first experimental tests of quantum mechanical uncertainty principles were provided by these systems \cite{Itano1993,Wineland1994}. Notably, through improvements in readout and control of single atoms \cite{Itano1993a,Diddams2001, Huntemann2016, Sanner2019}, and reduction of technical noise, the frequency precision of single ion atomic clocks has improved to reach a long-time frequency uncertainty of \cite{Itano1993, Ludlow2015}
\begin{equation}
\Delta \left(\hat{\omega} - \omega_0\right)^1 \approx \frac{1}{\sqrt{Tt}} > \frac{1}{T},
\label{eq:TF1}
\end{equation}
where $t < T$ is the phase evolution time of a single measurement, $\omega_0 = E/\hbar$ is the frequency of the atomic clock transition and $\omega$ is the frequency of the local oscillator to be estimated\footnote{Usually called Allen deviation, which is not exactly frequency uncertainty. A similar result for frequency uncertainty is obtained from the linewidth of a single measurement $= 1/t$, and taking multiple independent measurements up to time $T$. We are also being generous with the `$\approx$' usage here.}. The parenthesis indicates that $\omega$ is estimated with respect to known $\omega_0$ and the superscript explicitly defines the number of spin resources. Such a precision appears to be a fundamental limitation to atomic clock precision, since it is a result of the Heisenberg time-energy uncertainty principle \cite{Aharonov2002}.

Faced with hard limits on the absolute frequency uncertainty per unit time, the development of optical atomic clocks \cite{Ludlow2015} has enabled the fractional frequency uncertainty to be improved by increasing the carrier frequency. More recently, lattice clocks operating with $N = 100$ -- $1000$ ions \cite{Hinkley2013,McGrew2018} have allowed an uncertainty of
\begin{equation}
\Delta \left(\hat{\omega} -\omega_0\right)^N \approx \frac{1}{\sqrt{N Tt}} > \frac{1}{\sqrt{N} T},
\label{eq:TFN}
\end{equation}
to be achieved. In contrast to entangling proposals, interactions between atoms are explicitly avoided in lattice clocks since they shift the existing atomic energy transitions, introduce new transitions and cause decoherence, which reduce precision as the atoms are no longer independent.

As such, the current status is that, despite numerous proposals \cite{Wineland1992, Wineland1994, Kitagawa1993, Sanders1995, Bollinger1996, Ou1997, Zwierz2010, SchleierSmith2010, ElElla2021} and significant experimental effort \cite{Leibfried2004, Roos2006, Appel2009, Meyer2001, Leroux2010, LouchetChauvet2010}, entanglement has not been used to improve the frequency precision of any atomic clock beyond the SQL. Performance rather, has plateaued near this limit. Furthermore, there has been no satisfactory explanation as to why the experimental results do not attain the theoretical limits. A partial answer was provided by showing that in the presence of exponential decoherence, the maximally entangled state provides no improvement over the SQL \cite{Huelga1997} (see related work \cite{Fleischhauer2000} and \cite{Caves1985} in other contexts), but this does not explain the failure of entangling schemes in the presence of non-exponential decoherence or when optimal entangled states are used. Here we emphasize, that unlike in the field of quantum computation, achieving `quantum supremacy' in sensing does not require entanglement between large numbers of atoms or quantum error correction. The task is simple and apparently well within today's technology: use two atoms to demonstrate a frequency uncertainty of:
\begin{equation}
\Delta \left(\hat{\omega}-\omega_0\right)^2 < \frac{1}{\sqrt{2} T}.
\end{equation}
We are unaware of any experiment that has achieved this task.

Our review has focused on experimental results in the field of atomic clocks. This focus is not without reason. Control over quantum states in terms of readout fidelity, entangling fidelity, and decoherence rates is unsurpassed in trapped atoms and ions, thus one would expect the evidence for EEM to be strongest in these systems. More importantly, the metrological goal-posts are well defined and clearly reported in the literature, allowing state-of-the-art demonstrations to be compared with the theoretical limits. A more cynical explanation is that it is hard to be sneaky about atomic clock precision since, when presenting a frequency uncertainty as a function of measurement time, the results speak for themselves. 
Nevertheless, many of the papers we have cited actually claim to achieve an uncertainty below the SQL. For this reason, in \ref{App:Experiment} we review ten landmark papers that claim to demonstrate an uncertainty beyond the classical limits, providing a detailed analysis of the experimental uncertainty\footnote{We offer a bounty of US\$10,000 to anyone who provides a published article demonstrating an experimental uncertainty below Eq.\,\ref{eq:SQL}. Payable to the first article received. Send to \href{mailto:liam@gtoet.com}{liam@gtoet.com}. An additional bounty of US\$10,000 is offered to the first person to send a published article that demonstrates an experimental uncertainty below the $N = 1$ limit when using $N-$partite entanglement}. Effort is made to cover a broad field including magnetometry and optical phase measurements. We show that none of the experiments produce an uncertainty below that which can be obtained by employing the same resources without entanglement.

In summary, Eq.\,\eqref{eq:SQL} has limited the experimental uncertainty of a broad class of interferometric measurements in quantum metrology to date. This limitation has not been widely reported in the scientific literature and has arguably been actively misreported. If only one conclusion should be taken from this section, it is that investigation of this limit should be the focus of renewed experimental and theoretical effort.

\subsection{Theoretical principles in support of the central thesis}\label{Sec:Theory}
The previous sections have not introduced any theoretical analysis indicating that the SQL bounds the achievable measurement uncertainty, instead we have focused on outlining the underlying assumptions and the existence of experimental evidence that the HL can be attained. So what theoretical principles would lead one to reject the HL in favour of the SQL? This section proposes such a principle. In concert, we provide complementary analysis that excludes the HL from providing a strict experimental uncertainty bound.

Our argument begins by pointing out an inherent symmetry in atomic clock operation which assumes that estimation of the oscillator frequency with $N$ spins and $T$ time is limited by the uncertainty one can measure the energy of those $N$ spins in the same time (see Eq.\,\eqref{eq:TF1}). This duality can be exploited to derive general uncertainty limits for the estimation of any parameter $x$, by consideration of how $x$ shifts the atomic energy levels (\ref{App:SQL}). For time-keeping with atomic clocks where the detuning between an oscillator and the atomic frequency is measured, it can be summarized as follows \cite{Bollinger1996, Giovannetti2004, Aharonov2002}:

\begin{quotation}
\textit{Estimation of the oscillator frequency which drives an atomic transition is equivalent to estimating the transition frequency.}
\end{quotation}


We shall make prudential use of the symmetry between estimation of $\omega$ and $\omega_0$, however a vital component of the following analysis is the exploitation of asymmetries in the two estimation problems. Indeed we assert that blanket application of the above equivalency leads to derivation of incorrect uncertainty limits. The authority to make that statement comes from discovery of an estimation method which breaks the symmetry and which has been the key inspiration to formulation of this paper's the central thesis. In ref.\,\cite{McGuinness2021} we showed that it is possible to achieve $\Delta \hat{\omega}^1 \ll \frac{1}{N T}$, i.e. given only a single spin, the HL with any finite number of spins $N$ can be outperformed. Realization that Eq.\,\eqref{eq:HL} cannot bound the uncertainty for frequency estimation has in turn forced us to question the validity of the HL in general and particularly the dependence on $N$.

Specifically, one can develop a single atom estimation strategy which allows the frequency of a near-resonant field to be determined with an uncertainty bounded by\footnote{More precisely and using the notation of the previous section, we should write $\Delta \left(\hat{\omega}-\omega_c\right)^1$, where estimation of $\omega$ is performed with respect to the frequency of a known control field, $\omega_c$.}:
\begin{equation}
\Delta \hat{\omega}^1 > \sqrt{\frac{24 t_{\pi}}{T(T+t_{\pi})(2T+t_{\pi})}},
\label{eq:qdyne}
\end{equation}
where $t_{\pi}$ is the time needed to perform a $\pi$-rotation on the spin sensor \cite{McGuinness2021}. Apart from a non-linear time dependence going below $1/T$ for $T \gg t_{\pi}$, the bound differs from other uncertainty bounds in quantum metrology in several other characteristics. In particular the bound:
\begin{itemize}
\item is not obtained from estimating the energy of the spin sensor.
\item is insensitive to spin decoherence and the coherence time of the sensor.
\item is optimized when more than one measurement (actually as many as possible) is performed during $T$.
\item does not require a small prior uncertainty because the bandwidth does not decrease with sensitivity or $T$.
\end{itemize}
Eq.\,\eqref{eq:qdyne} is derived by considering a dataset obtained in time $T$, from multiple measurements on a single atom, subject to the Hamiltonian:

\begin{align}
H(t_0,t/2) &=\frac{\hbar \omega_0}{2}\sigma_z + \frac{\hbar \Omega}{2} \left( \cos \left[ \omega t/2 + \varphi(t_0) \right] \sigma_x + \sin \left[ \omega t/2 + \varphi(t_0) \right] \sigma_y \right) \nonumber \\
H(t/2,t) &=\frac{\hbar \omega_0}{2}\sigma_z +  \frac{\hbar \Omega}{2} \left( \cos \left[ \omega_c t/2 \right] \sigma_x + \sin \left[ \omega_c t/2 \right] \sigma_y \right), \label{eq:Ham}
\end{align}
where $t_0$ denotes the measurement start time, $t$ is the phase evolution time of a single measurement, $\omega_{c}$ is the frequency of a control field with known frequency, $\Omega$ the Rabi frequency of the atom-field interaction, $\varphi(t_0)$ the phase of the signal field at $t_0$, and $\sigma_{x,y,z}$ the Pauli spin-matrices. Generating the Hamiltonian requires that the experimentalist has control over the amplitude of the signal field and the phase (frequency) and amplitude of the control field. Setting $t = \pi/\Omega \equiv t_{\pi}$ the two fields sequentially interact with the spin in order to perform two $\pi/2$-rotations around axes defined by their phases. Measurement of the spin population along the $z$-axis then provides information on the relative phase difference between the \emph{control} and \emph{signal} fields, and not the relative difference between the atomic state and the signal phase.

The idea behind the technique is to use many measurements of the phase difference per unit time between the signal and control fields to determine their relative detuning; as opposed to the detuning between the signal frequency and the atomic Larmor. As the atomic frequency is not estimated, temporal correlations in the measurement estimator\footnote{The estimator is $|\hat{\varphi}(t_i)-\varphi_c (t_i)| \Rightarrow |\hat{\omega}-\omega_{c}|$, where $t_i$ is the midpoint of the $i^{\mathrm{th}}$ measurement and we have the freedom to set $\varphi_c (t_i) = 0$.} remain despite multiple measurements being performed on the spin sensor, resulting in a dataset that resembles a heterodyne measurement \cite{Schmitt2017, McGuinness2021, Boss2017, Meinel2020}. The frequency uncertainty from heterodyning is known to surpass $1/T$ when `classical' noise is considered and a detector comprised of many atoms is used \cite{Kay1993}, and this property is preserved when quantum projection noise from readout of a single atomic sensor is considered, leading to Eq.\,\eqref{eq:qdyne} above.

\begin{figure}[thb]
\includegraphics[width=0.48 \textwidth]{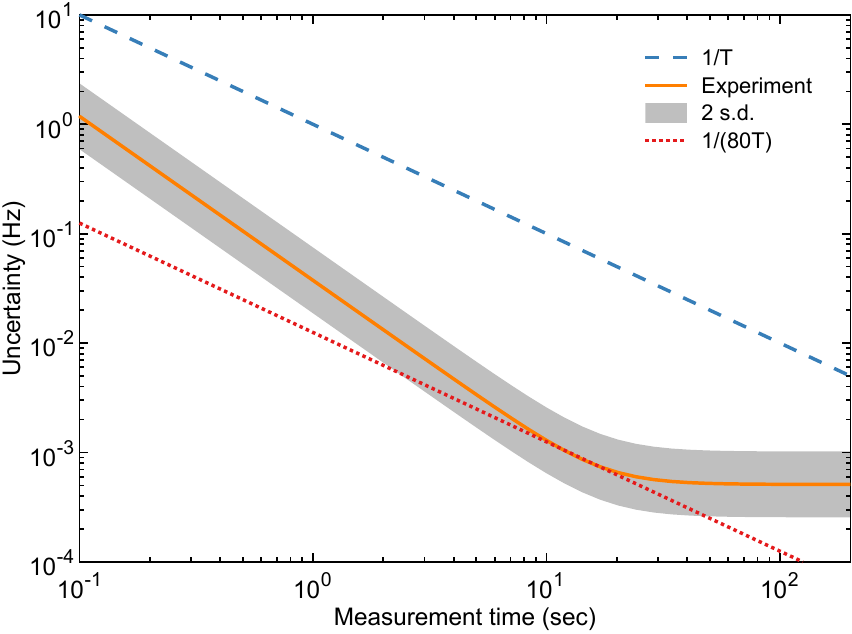}
\protect\caption{\textbf{Frequency estimation with a single spin}. Experimentally reported frequency uncertainty as a function of measurement time for estimation of a 5 MHz radio-frequency field with a single nuclear spin (orange line) and two standard deviation confidence region (grey). Comparison to $1/T$ and $1/(80T)$ functions indicated by blue dashed and red dotted lines respectively. Reproduced from the results in \cite{McGuinness2021}.}
\label{fig1}
\end{figure}

As noted for the standard quantum and Heisenberg limits, it is also clear that Eq.\,\eqref{eq:qdyne} cannot be saturated due to approximations -- e.g. instantaneous readout and initialization with unit fidelity -- in the derivation of the lower bound. Furthermore, the uncertainty scaling of $T^{-3/2}$ cannot continue indefinitely, since this would require $\omega_{c}$ to be known perfectly. The pertinent question is whether this is a loose bound, as we assert for the HL, or whether the bound is approachable? Importantly, the frequency estimation technique has been experimentally demonstrated, providing an achievable limit \cite{McGuinness2021}. Considering non-unity readout and initialisation fidelity of a single spin taking $\sim 1\,\mu$s, an uncertainty within a factor 5 of the theoretical bound was obtained, which persisted for several tens of seconds until systematic uncertainty in $\hat{\omega}_c$ dominated. The experimental uncertainty is well described by:
\begin{equation}
\left( \Delta \hat{\omega}_{\mathrm{exp}}^1 \right)^2 = \left(C \Delta \hat{\omega} ^1\right)^2 +\left( \Delta \hat{\omega}_c \right)^2
\end{equation}
where $C$ depends on the readout and initialisation parameters and $\Delta \hat{\omega}_c$ is the systematic uncertainty in the control frequency. The experimental values reported in ref.\,\cite{McGuinness2021}, with $C=4.7$, $t_{\pi} = 5 \times 10^{-6}$\,s give:
\begin{equation}
\left( \Delta \hat{\omega}_{\mathrm{exp}}^1 \right)^2 = \frac{1.4 \times 10^{-3}}{T^3+8 \times 10^{-6}T^2+1.4 \times 10^{-11}T} +\left(5.1\times 10^{-4}\,\mathrm{Hz}\right)^2,
\label{eq:qdyneexp}
\end{equation}
going more than 10 standard deviations and nearly two orders of magnitude below the $1/T$ uncertainty limit (Fig.\,\ref{fig1}).
\\

We have taken the time to discuss this experiment in-depth because the results rule out the oft stated Heisenberg limit providing the lower bound for frequency estimation with $N$ atoms:
\begin{equation}
\Delta \hat{\omega}_{\mathrm{HL}}^N > 1/NT.
\label{eq:TF}
\end{equation}
Eq.\,\eqref{eq:TF} cannot bound the uncertainty since a lower uncertainty for $N < 80$  has been demonstrated with access to only a single atom. In expectation of resistance to this assertion, several objections are addressed in \ref{App:objections}, the strongest being that we have implicitly made use of extra resources not considered in the derivation of Eq.\,\eqref{eq:TF}. We argue strongly that this is not the case.

Eq.\,\eqref{eq:qdyne} does not, in itself, contradict the SQL or HL for estimation of $x$, since $\omega$ is not a multiplicative parameter in the Hamiltonian and is not in the same class of parameters as $x$ for which the limits are derived\footnote{By changing to a frame rotating at either the Larmor or field frequency, the detuning $|\omega - \omega_0|$ appears as a multiplicative parameter in the Hamiltonian and thus estimation of this detuning is limited by Eq.\,\ref{eq:SQL}. The solution we present involves interaction with two fields of different frequencies, thus there is no single rotating frame, the detuning is not a multiplicative parameter and the equivalency leading to Eq.\,\ref{eq:SQL} does not hold.}. The lack of recognition of this difference in the quantum metrology literature has lead to derivation of incorrect uncertainty limits. It's significance, however, is in providing a counter-example to a central assumption of quantum metrology; that the number of sensor particles has an equivalent metrological power to sensing time. This observation should instil a sense of uncertainty into acolytes of the Heisenberg limit, especially since the duality between estimation of either the oscillator or the atomic frequency in atomic clocks means that a non-linear time dependence for one problem is a strong impetus to question assumptions leading to the HL in the other scenario. 

Via the following sequence of logical arguments, we argue that the HL cannot provide a strict bound in general, where additional details on each of the steps is provided immediately below.
\begin{enumerate}
\item Eq.\,\eqref{eq:qdyne} prohibits a linear uncertainty scaling in $N$ for estimation of $\omega$.
\item The HL prohibits a better than linear scaling in $N$ for estimation of $\omega_0$.
\item The estimation limits for $\omega_0$ and $\omega$ coincide for (at least part of) a single measurement.
\end{enumerate}

Beyond ruling out the HL on bounding the uncertainty for frequency estimation, the form of Eq.\,\eqref{eq:qdyne} indicates that the bound cannot be linear in atom number $N$, since reaching an uncertainty of: 
\begin{equation}
\Delta \hat{\omega}^N \approx \frac{1}{N} \sqrt{\frac{24 t_{\pi}}{T(T+t_{\pi})(2T+t_{\pi})}},
\label{eq:wrongqdyneN}
\end{equation}
would require not only that the readout and initialisation time and fidelity are unaffected by entangling/disentangling protocols but also that the $\pi$-rotation time, $t_{\pi}$ remains the same for an entangled state, which is in conflict with the Margolus-Levitin \cite{Margolus1998} and Mandelstam-Tamm bounds \cite{Mandelstam1991}. In fact, a speed-up in state evolution (to an orthogonal state) is the basis behind proposed precision enhancements with entangled states, thus we would expect $t_{\pi}$ to also be reduced by a factor of $N$, resulting in a precision of:
\begin{equation}
\Delta \hat{\omega}^N \approx \frac{1}{N} \sqrt{\frac{24 t_{\pi}/N}{T(T+t_{\pi}/N)(2T+t_{\pi}/N)}}.
\label{eq:wrongqdyneN2}
\end{equation}

Such a modification preserves the symmetry between sensor number and sensing time. But now, in order to consistently maintain that entanglement provides any metrological enhancement, we are forced to accept that, for frequency estimation, entanglement provides a super-linear enhancement with sensor number. Moreso, at short times, the frequency uncertainty reduces as $T^{-2}$ for a single measurement \cite{Pang2017, Schmitt2021}, thus demanding quadratic improvement with sensor number to preserve symmetry. Strict adherents to entanglement advantages may maintain that this is indeed the correct limit, but then they should also note that such theoretical predictions diverge markedly from empirical reality. No improvement over $\sqrt{N}$ has been demonstrated for any estimation task when time is considered, let alone quadratic improvement. Experimental discrepancies aside, there are theoretical issues with allowing a greater than $N$ uncertainty improvement for estimation of $\omega$. Aharanov, Massar and Popescu have proved that the uncertainty is bounded by $\Delta \hat{\omega} > 1/(4NT)$ for a single measurement, assuming that the energy scales linearly with $N$.

\begin{figure}[bt]
\includegraphics[width=0.95 \textwidth]{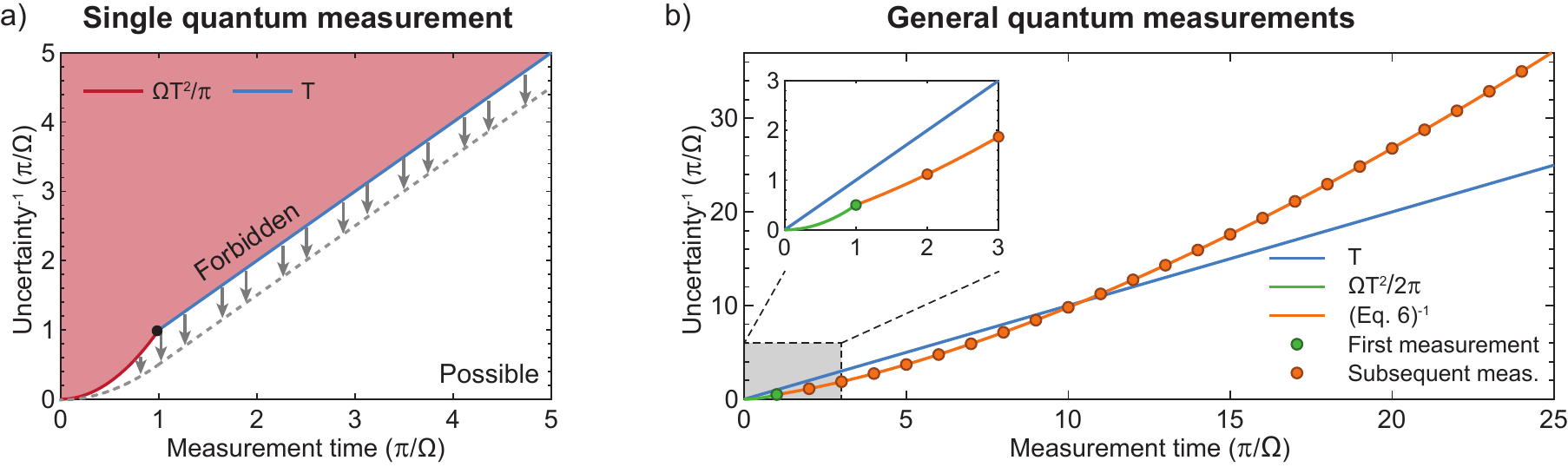}
\protect\caption{\textbf{Frequency uncertainty of quantum measurements.} (a) For any single quantum measurement, the frequency uncertainty is bounded by $\pi/(\Omega T^2)$ (red) for $T < \pi/\Omega$ (black circle) and thereafter bounded by $1/T$ (blue). Stricter analysis limits the uncertainty to $2\pi/(\Omega T^2)$ for $T$ up to $\pi/\Omega$ (\ref{App:Experiment}) with a linear improvement thereafter (dotted line). (b) A frequency uncertainty beyond the $1/T$ limit results from the scaling of multiple measurements when performing heterodyning. For a single sensor, the first measurement has an uncertainty bounded by $2 \pi/(\Omega T^2)$ (green) for $T < \pi/\Omega$. The uncertainty obtained from subsequent measurements is bounded by Eq.\,\ref{eq:qdyne} (orange). No individual measurement provides an uncertainty less than $1/T$ (blue), but the uncertainty from a dataset composed of many measurements, performed at times marked by filled circles, surpasses this bound for $T > 10 \pi/\Omega$. (Inset) Uncertainty improvement for the first three heterodyne measurements and comparison to $1/T$ bound. $N = 1$ in all figures.}
\label{fig2}
\end{figure}

Figure\,\ref{fig2} illustrates how the different temporal scaling of Eq.\,\ref{eq:qdyne} can be reconciled with the bound of Aharanov, Massar and Popescu. In Fig.\,\ref{fig2}a) we show for a measurement time less than $\pi/\Omega$, where $\Omega$ is the intensity of the signal and control fields, the frequency uncertainty improves quadratically and remains worse than $1/T$ (red line) \cite{Pang2017, Schmitt2021}. Thereafter the uncertainty improves linearly with time (blue line) at which point (marked with a black circle) it is advantageous to readout the sensor and estimate $\omega$ if performing a heterodyne measurement, since additional measurements provide super-linear estimation reduction, assuming the signal is not destroyed by the measurement. Alternatively, if one is estimating $\omega_0$ or $\omega$ via Ramsey interferometry, it is preferable to wait longer and take advantage of the linear uncertainty reduction afforded by a single measurement before reading out the sensor.

Figure\,\ref{fig2}b) shows that the improved estimation obtained with heterodyne arises from the information provided by multiple measurements, and that taken in isolation, the first (or any other individual) measurement does not outperform Ramsey interferometry and the $1/T$ limit. Indeed, we have that the uncertainty improvement with homodyne (e.g. a Ramsey measurement) or heterodyne (e.g. a Qdyne measurement as implemented in ref\,\cite{McGuinness2021}) estimation is initially identical since they both start with a $\pi/2$-rotation induced by the signal field. The key point is that for (at least part of) a single measurement, the uncertainty improvement with $T$ must be identical for homodyne and heterodyne measurements, as the measurement protocols themselves are identical. We then require by symmetry that within a single measurement, the uncertainty improvement with $N$ is also identical for both schemes.

We have thus shown that both measurement protocols are identical for a time during which, given assumptions on symmetry, a linear improvement or a super-linear improvement with $N$ is excluded. I.e. analysis of the uncertainty improvement with $N$ using a heterodyne protocol prohibits a linear improvement whereas a super-linear improvement is prohibited with a homodyne protocol. Thus we are left with $\sqrt{N}$ improvement being the best remaining candidate. Furthermore, the $T^{-3/2}$ uncertainty of the heterodyne measurement is experimental proof that Ramsey interferometry must have an initial $T^{-2}$ uncertainty scaling, since this is required in order to observe the uncertainty reduction observed with heterodyne detection. A direct implication is that perfect and instantaneous readout in any measurement basis is not possible, otherwise a linear improvement could be achieved with Ramsey interferometry at all timescales.

Ref.\,\cite{McGuinness2021} provides an additional theoretical argument that excludes the uncertainty from entangling schemes from outperforming independent particles. As entanglement often requires temporal overheads for entanglement (and disentanglement) generation, and since no signal is acquired during entanglement generation, then the uncertainty from such schemes will initially be outperformed by independent particles. Proponents of EEM are therefore arguing that the lowest uncertainty path through the metric of quantum states is not continuous. In effect, that it is possible for the state which achieves optimal uncertainty at some time to be provably non-optimal at some earlier time when the overheads are implemented, and that no continuous transform exists that takes the optimal state at an earlier time to the optimal state at a later time. The principle of continuous optimal sensitivity rules out such entangling schemes by allowing one to determine the optimal measurement sequence without requiring prior knowledge of the available measurement time.

In addition to providing theoretical arguments in support of the central thesis, we now provide a physical principle from which it can be derived. We argue that the entire debate on the possibility of EEM can be reduced to a single conjecture; that no individual measurement on a single system can provide a frequency uncertainty below $1/T$, where $T$ is the measurement time (see ref.\,\cite{McGuinness2021} and the 1TF conjecture within). Here, two or more entangled atoms are to be treated as a single system since this is the defining feature of entanglement - that entangled atoms cannot be described separately. We argue that the conjecture is likely correct, and is physically motivated since it coincides with the Gabor-Fourier limit for waveform estimation \cite{Gabor1947, Cohen1995}. 
With this conjecture, one can show that entanglement cannot provide a frequency uncertainty below \cite{McGuinness2021}:
\begin{equation}
\Delta \hat{\omega}_{\mathrm{SQL}}^N > \sqrt{\frac{24 t_{\pi}}{NT(T+t_{\pi})(2T+t_{\pi})}},
\label{eq:qdyneN}
\end{equation}
which is a $\sqrt{N}$ improvement on the uncertainty with a single sensor.

From this limit, relationships to other estimation strategies can be exploited in order to tighten the estimation bound beyond the general HL. Most directly, acceptance of Eq.\,\eqref{eq:qdyneN} requires that we must modify at least one of our beliefs regarding the possibility to estimate the transition frequency of $N$ atoms in time $T$ according to:
\begin{equation}
\Delta \hat{\omega}_{0,\mathrm{HL}}^N \approx 1/NT.
\label{eq:TF0}
\end{equation}
Either:
\begin{enumerate}
\item[] $\omega_0$ can be estimated independent of knowledge on $\omega$. I.e. there exist strategies that allow one to estimate the transition frequency of an atom without frequency dependent absorption or emission of a photon. This is depicted as the grey square in Figure\,\ref{fig2}a).
\end{enumerate}
Or
\begin{enumerate}
\item[] Eq.\,\eqref{eq:TF0} does not hold in general, since if $\Delta \left(\hat{\omega}_0 - \omega \right) = \frac{1}{NT} + \epsilon$, for small constant $\epsilon$, then by symmetry this would imply a violation of Eq.\,\eqref{eq:qdyneN}.
\end{enumerate}

In presenting these two options, we have rejected several other possibilities. A third option is to require that Eq.\,\eqref{eq:TF0} holds only for time longer than $T_{\mathrm{min}}$ in order to prevent conflict with Eq.\,\eqref{eq:qdyneN}:
\begin{equation}
T_{\mathrm{min}} = \frac{t_{\pi}}{4}\left(-3 + 24N + \sqrt{1-144N+576N^2}\right).
\end{equation}
Meaning that when entangling the atoms, obtaining a measurement uncertainty of $\Delta \hat{\omega}_0 \approx 1/NT$ must take at least $T_{\mathrm{min}}$ time. But as $T_{\mathrm{min}}$ seems completely arbitrary (for $N = 2$ we have $T_{\mathrm{min}}\approx 22.5 t_{\pi}$) and does not depend on the entangling, readout or initialization rate, we believe it cannot be correct.

Alternatively, one could allow that information on $\omega_0$ be obtained via interaction with a field oscillating with frequency $\omega \simeq N \omega_0$. Indeed, this can be achieved by bringing the spins close together so that for all practical purposes they form a single spin with $N$ times larger magnetic moment (but not by entangling the spins). Measurement of the relative detuning: $\Delta (\hat{\omega} - N \omega_0 ) \approx 1/T$ then allows estimation of $\omega_0$ at the Heisenberg limit without violation of $\Delta \hat{\omega} > 1/T$. However to do this, we have replaced $N$ spins with a single spin, so the uncertainty dependence on sensor number is lost. This is exactly why entanglement has been proposed to overcome the loss of resource number, since there is a well-defined sense in which each atom can be readout and controlled individually, even if the measurement results are correlated between atoms. However, when explicit entangling protocols are presented \cite{Wineland1992, Wineland1994, Bollinger1996, Meyer2001, Leibfried2004, Appel2009, Gross2010, Leroux2010, LouchetChauvet2010, Bohnet2016, Pezze2018, Shaniv2018}, each atom still interacts with a field of frequency $\omega \simeq \omega_0$, so this route appears closed.

Given the choices outlined above, and without a method to measure the atomic frequency without interrogating the frequency of a classical oscillator, we advocate that the standard quantum limit provides a bound on $\Delta\hat{\omega}_0$. This in turn limits the precision one can estimate any general parameter $x$, such as the external magnetic field, satisfying the properties leading to Eq.\,\eqref{eq:SQL}.

\emph{A final remark}: The 1TF conjecture implies that not only does $\Delta \hat{x} > \frac{\lambda}{\sqrt{N}T}$ constitute a fundamental uncertainty limit which cannot be surpassed, it also provides a more stringent uncertainty bound on entangled systems. Namely, if full entanglement between $N$ sensors is exploited, then the uncertainty in parameter estimation is bounded by:
\begin{equation}
\Delta \hat{x} > \lambda/T,
\end{equation}
i.e. the entangled system will not outperform the optimum precision that can be obtained when using just one of the $N$ atoms. This statement is much stronger than the central thesis alone, and subsequently bears more divergent predictions. In the literature review presented in \ref{App:Experiment} we make note of this single spin estimation limit and show that the current experimental evidence is consistent with the predictions. With irony we note this would mean that the expenditure of money and resources in the hope of achieving a $\sqrt{N}$ entanglement improvement over classical methods has actually gone towards reducing performance. How much worse, one might ask? By precisely a factor of $\sqrt{N}$.


\subsection*{Conclusion and outlook}
By invoking basic physical/informational principles and assessing the currently available experimental evidence, we have put forward the case that entanglement cannot improve the performance of quantum sensors beyond that ultimately achievable with independent sensors. These claims have strong predictive powers which can be experimentally tested:
\begin{enumerate}
\item No sensing device containing $N$ atoms can be constructed that achieves a precision $\sqrt{N}$ below that ultimately achievable with a single atom in the same amount of time.
\item No metrological device fully exploiting entanglement can be constructed where the precision exceeds that achievable with a single atom or photon in the same amount of time.
\end{enumerate}

Beyond quantum metrology, this work impacts several related areas. Childs, Preskill, and Renes have formalised connections between quantum sensing and quantum computation that allow one to bound the performance in information processing from the uncertainty in precision sensing \cite{Childs2000} . Therefore we expect that methods similar to those introduced here can be used to show quantum computers will not outperform classical computers for general information processing tasks. In addition, it is often stated that, due to remarkable agreement with experiments, quantum mechanics is our most accurate description of nature (see e.g. Ch.\,2 of \cite{Nielsen2000}). Here, we have outlined significant unreported discrepancies between experimental results and standard interpretations of quantum mechanics, in particular an $N$-fold discrepancy between experimental uncertainties and the Heisenberg uncertainty limit with entanglement.

These observations provide two perspectives to modifying quantum mechanics. Firstly, intrinsic non-linearities must be incorporated in quantum mechanical descriptions to move away from a linear uncertainty reduction with time and atom number. Secondly, saturation of the Heisenberg limit assumes ``in line with Neils Bohr's interpretation of non-relativistic quantum theory -- that in an instantaneous measurement any observable, by itself, can be measured arbitrarily accurately ..." (Caves \emph{et. al.} 1980 \cite{Caves1980a}). Our assertion that the Heisenberg limit is not achievable therefore requires modifying the assumption of instantaneous measurements with arbitrary accuracy. This is precisely where we propose that non-linearities should enter into quantum theory. Although perhaps a mundane modification of quantum mechanics, it nevertheless conflicts with the standard application of quantum theory -- the Copenhagen interpretation.

Finally, we note that the ideas provided here are readily comprehensible to an undergraduate student; indeed a case can be made that the presented logic and physical arguments are more intuitive than the current dogma. Furthermore, the experimental evidence is obvious to anyone who looks. So why has the standard narrative persisted unchallenged? Perhaps the sociological and psychological lessons from addressing that question will provide greater impact than the physical statements provided here (irrespective of whether they turn out to be correct).

\bibliographystyle{naturemag}
\bibliography{references2.bib}

\renewcommand*{\thesection}{SI}
\setcounter{subsection}{0}

\newpage
\setcounter{page}{0}

\section*{Supplementary Information for The case against entanglement improved measurement precision}
\subsection{Derivation of the standard quantum limit: Eq.\,\eqref{eq:SQL}}\label{App:SQL}
The standard quantum limit (SQL) bounds the uncertainty in estimating an unknown classical parameter $x$, with $N$ independent and identical resources (taken to be $S = 1/2$ spins here) in time $T$. A non-standard derivation of the SQL starts by assuming the following Heisenberg time-energy (T-E) uncertainty relation:
\begin{equation}
\Delta\hat{E}^1 > \frac{\hbar}{T},
\label{eq:TE}
\end{equation}
to be understood as --- the uncertainty in estimating the transition energy $E$, of a single spin must be at least the reduced Planck constant $\hbar$, over the measurement time, where the superscript explicitly defines the number of spins to be estimated. If the spin interaction with parameter $x$ can be written as $xH$ for some time-independent Hamiltonian $H$, then we can derive the following uncertainty relation:
\begin{equation}
\Delta\hat{E}(x)^1 > \frac{\hbar}{T} \quad \Rightarrow \quad \Delta\left(\hat{x} \frac{\hbar}{\lambda} - E_0 \right)^1 > \frac{\hbar}{T} \quad \Rightarrow \quad \Delta\left(\hat{x} - \lambda \omega_0  \right)^1 > \frac{\lambda}{T},
\label{eq:deltax}
\end{equation}
where $E_0 = \hbar \omega_0$ is the interaction-free transition energy with transition frequency $\omega_0$, $\lambda$ converts the units of $x$ to time and the parenthesis indicates that $x$ is estimated with respect to $\lambda \omega_0$. We have that $\Delta\hat{x}$ is minimized when $\Delta \left(\lambda \omega_0 \right) = 0$, assuming $\Delta\hat{x}$, $\Delta \left(\lambda \omega_0 \right)$ are independent. Restricting ourselves to measurements on a single particle, Aharonov, Massar, and Popescu \cite{Aharonov2002}, claimed a proof of this relation with a factor of 1/4 lower bound, adding that ``we believe that this lower bound is not tight and that in general a stronger lower bound should hold". The stronger bound that they refer to coincides with Eq.\,\ref{eq:TE}. Their proof operates in an inverted direction, where first an uncertainty bound in estimating an unknown Hamiltonian is derived and that result is used to bound the uncertainty in energy estimation. However Aharonov, Massar, and Popescu only bound the uncertainty in estimating the Hamiltonian from a single measurement, with the implicit assumption that this saturates the bound. In ref. \cite{McGuinness2021} we have shown how the bound can be surpassed by performing multiple measurements, thus violating their proof. For estimating the energy of a single spin, the assumption of saturation from a single measurement is believed to hold, thus their proof is valid for this estimation task. However, this assumption is equivalent to the T-E uncertainty principle, so we find our starting point to be equally justified without needing to impose restrictions on the number of measurements that can be performed.

Assuming that we perform identical and independent measurements on $N$ identical and independent spins, the SQL follows:
\begin{equation}
\Delta \hat{x}_{\mathrm{SQL}}^N > \frac{\lambda}{\sqrt{N}T}.
\label{eq:SQLapp}
\end{equation}

Note, we have not needed to invoke any fundamental noise source or speculate on the achievable signal-to-noise with which we can measure the spins. The SQL was derived solely from the T-E uncertainty relation, where we associated $\hat{x}$ with estimation of the spin transition energy $E$, provided several assumptions. This is the power of our approach. We now address the underlying assumptions.

\subsubsection{Parameter class x}
The SQL that we have derived only holds for parameters that can be written as a multiplicative constant in the spin Hamiltonian. We also assume that the Hamiltonian is time-independent, although we suspect that this caveat can be relaxed so long as the energy remains fixed during the experiment (see ref.\,\cite{Aharonov2002}). The first substitution in Eq.\,\ref{eq:deltax} assumes that $xH$ is diagonalizable in the same basis as $H$ to maximise information via the Cauchy-Schwartz inequality. We also require that $x$ has a fixed and finite value, in both time and space, so that each spin sees the same value of $x$ at all times. We assume that more atoms can be used to measure $x$ and also more than one measurement can be performed without the value of $x$ changing. In this respect $x$ is a classical parameter. Relaxing this assumption, so that the parameter to be estimated contains a finite number of atoms or photons, results in worse estimation bounds.

Describing the frequency of a near-resonant field as a classical parameter $\omega$, the uncertainty bounds for this estimation problem as derived in Eq,\ref{eq:qdyneN} conflict with the SQL (Eq.\,\ref{eq:SQLapp}). We obtain a different uncertainty limit for this estimation problem because the preceding assumptions are not satisfied, since $\omega$ does not appear as a multiplicative constant in a time-independent spin-interaction Hamiltonian.


\subsubsection{Estimation of the spin transition energy}
Our derivation of the SQL assumes that the value of $x$ is inferred from estimation of the spin transition energy. If we accept that the T-E relation limits the estimation uncertainty for a single spin, then for parameters in class $x$ we can show this is an efficient estimation strategy. To see this note that information on $x$ is obtained by measuring a change in spin state, dependent on $x$, from some initial state $\ket{\Psi (0)}$ to some final state $\ket{\Psi (t)}$. The actual metric being provided by $\cos^{-1}\left[|\braket{\Psi (t)}{\Psi (0)}|\right]$ \cite{Wootters1981, Braunstein1994}. From the Schr\"odinger equation we have $\ket{\Psi (t)} = e^{-i xH t /\hbar}\ket{\Psi(0)}$, giving $\frac{\partial \ket{\Psi (t)}}{ \partial x} = -i E t \ket{\Psi (t)}$, thus the sensitivity to $x$ depends on the spin energy and time, just as we have obtained by the T-E uncertainty relation. However, here, $t$ is the coherent evolution time and does not include time preparing or reading out the spin, thus saturation of the SQL requires perfect and instantaneous readout.

However, for general parameter classes, estimation of the spin transition energy is not always optimal, and for different interaction Hamiltonians more efficient estimation strategies exist. It turns out that instead of estimating the spin eigenvalues, we can estimate the eigenvectors of the Hamiltonian and obtain different uncertainty limits. For frequency estimation, this type of strategy surpasses the T-E uncertainty principal and results in the uncertainty bound of Eq,\ref{eq:qdyne}.

\subsubsection{Prior information}
What do we mean by `unknown' when we say estimation of an unknown parameter $x$? If the range of $x$ is finite, e.g. a phase measurement where $0 < x \le 2 \pi$, then unknown could mean that the prior probability of $x$ is uniform across the range and we do not require any prior information on $x$. However, even for parameters with finite range, saturation of the uncertainty bounds generally only occurs for a small range of parameter space, thus prior information is required to reach optimal uncertainty. But what about parameters with unbounded range? For estimation of unbounded parameters with finite resources we conjecture that prior information in the form of a finite \emph{a priori} uncertainty is required in order to obtain a non-zero bound.
Then the uncertainty of $\hat{x}$ resulting from one measurement of a single spin-1/2 system, taking time $T$ is bounded by:
\begin{align}
\Delta \hat{x} > \begin{cases}
\lambda /T_0, & T < T_0 \\
\lambda /T, & T \geq T_0
\end{cases}
\label{eq:prior}
\end{align}
where the \emph{a priori} uncertainty of $x$ is $\Delta\hat{x}^0 = 2 \lambda/T_0$. The bound of Eq.\,\ref{eq:prior} has two parts. State readout of a single qubit cannot provide more than 1 bit of information, which in turn cannot reduce the uncertainty more than a factor of two below the prior \cite{Shannon1948, Peres1991}. This is just Shannon classical information theory predicated on state readout of a single atom providing one bit of information. Indeed the Heisenberg uncertainty relations for non-commuting observables prevent more than one bit of information to be extracted. The second part of the bound states that there is a finite amount of information per unit time that can be extracted from a single spin. This allows that there are non-state readout estimation strategies such as recording the frequency of the emitted photon with a spectrometer that allow more than one bit of information to be extracted per measurement, but that these strategies do not provide better than $1/T$ improvement. Allowing for multiple state readout measurements, we also require a bound on reduction of the prior by more than a factor of two, since the posterior of one measurement becomes the new prior, the timescale for this uncertainty reduction must be bounded so that the T-E relation is not violated. Our derivation of the SQL relates to the second case of Eq.\,\ref{eq:prior}, where we assume that the prior information is less than $2 \lambda/T$.

\subsubsection{State preparation}
In order to avoid that the $N$ spins have been prepared with some unaccounted resources, we place restrictions on the initial state of the spins which will be inserted into the Ramsey interferometer. In particular we require that the initial state of the spins is independent of $x$. 
Actually, the most natural requirement is that the spins should be in a fully mixed state at the start of any measurement. In effect we demand that we are given the spins without knowledge of the state they are in, which ensures that any information extracted cannot have come from their preparation. Maybe one can argue that this is being unnecessarily restrictive, but if entanglement enhancements can only be obtained from assumptions on state preparation, then this should be made clear, and it also doesn't seem to fit with most peoples expectations on the power of entanglement. Furthermore, if the assumption of perfect and instantaneous measurements in any basis holds, then this requirement does not in fact place any restrictions, since the initial mixed state can be measured and purified instantaneously without any overhead.

\subsection{Modification of the SQL}\label{App:1}
We add two corollaries to the central thesis that modify the uncertainty bound of the SQL at each time extremum. First, after some finite time $t_{\times}$, we assert that the uncertainty is bounded by:
\begin{equation}
\Delta\hat{x} > \frac{\lambda}{\sqrt{N T t_{\times}}}, \quad  \left\lbrace T > t_{\times} \right\rbrace
\label{eq:Squareroot}
\end{equation}
even for a single measurement. This bound just states that there is a physical reason, such as decoherence or imperfect prior knowledge which prevents the uncertainty from improving linearly for an indefinite time.

Second, up to some finite time $t_{\star}$, we assert that the uncertainty is bounded by:
\begin{equation}
\Delta\hat{x} > \frac{\lambda t_{\star}}{\sqrt{N} T^2 }, \quad  \left\lbrace 0 < T \leq t_{\star}\right\rbrace,
\label{eq:quadratic}
\end{equation}
i.e. below some short time, the uncertainty improves non-linearly and remains below the $1/T$ limit. The timescale should be related to the maximal bit-rate that one can readout a quantum system, so $t_{\star}$ represents the minimum time to obtain one bit of information. In order to obtain a full bit of information on $x$, it must at least in principle, require enough time for the quantum system be driven to an orthogonal state, therefore $t_{\star}$ must be at least $t_{\pi}$, the minimum time to perform a $\pi$-pulse on the quantum system.

The bound of SQL (Eq.\,\ref{eq:SQL}) is also a loose bound in that it assumes instantaneous and perfect fidelity readout (and initialisation) of the spin sensor which cannot be realised. By taking into account the amount of time required to achieve a given readout fidelity (which much be less than unity) further restrictions on the SQL can be placed. However this requires careful characterisation of the physical resources implemented in readout the quantum sensor, which is beyond the scope of this paper. Finally, given that $x$ is a physical parameter, we also have that $\sqrt{N}$ improvement cannot continue indefinitely for an arbitrarily large number of sensors, since it would require $x$ to contain infinite resources and be of infinite spatial extent.

\subsection{Review of experimental evidence against the SQL}\label{App:Experiment}
We review several widely cited papers that, at face value give the impression of exceeding the SQL. Common loopholes appear so regularly in the literature that we define each one below in order to refer to them during the critique. The loophole strategies are:

\begin{enumerate}
\item \textbf{Forget time:} An uncertainty of $\Delta \hat{x} = \frac{\alpha \lambda}{N}$ is achieved for $1 < \alpha < \sqrt{N}$ which results in $\Delta \hat{x} < \frac{\lambda}{\sqrt{N}}$. However, when the time to achieve this uncertainty is included in the analysis, the result is $\Delta \hat{x} > \frac{\lambda}{\sqrt{N}T}$, i.e. worse than the SQL. Sometimes just some of the time is included in the analysis -- the phase accumulation time -- so that $\frac{\lambda}{\sqrt{N} T} < \Delta \hat{x} < \frac{\lambda}{\sqrt{N} t}$. \label{item_time}, which is also bounded by the SQL. \\

\item \textbf{Multiple independent measurements:}  An uncertainty of $\Delta \hat{x} = \frac{\alpha \lambda}{N\sqrt{T}}$ is achieved for $1 < \alpha < \sqrt{N}$ which results in $\Delta \hat{x} < \frac{\lambda}{\sqrt{N T}}$. In this analysis it is assumed that multiple identical and independent measurements must be performed in time $T$, leading to an uncertainty improving only as $1/\sqrt{T}$. When compared to linear time improvement of the SQL, the uncertainty does not go below this bound. \label{item_multiple} \\

\item \textbf{Scaling isn't everything:} An uncertainty scaling of $\Delta \hat{x} = \frac{\alpha \lambda}{N^{\beta}T}$ is achieved for $\beta >\frac{1}{2}$, which scales better than SQL. However, $\alpha > N^{\beta - 1/2}$, so the absolute value of the uncertainty is $\Delta \hat{x} > \frac{\lambda}{\sqrt{N}T}$. An enhanced scaling in terms of time, may also be achieved: $\Delta \hat{x} = \frac{\alpha \lambda}{\sqrt{N} T^{\beta}}$ for $\beta > 1$. However in this case $\alpha > T^{\beta - 1}$ so the absolute uncertainty does not go below the SQL. \label{item_scaling} \\

\item \textbf{Imperfect isn't the limit:} The experimental uncertainty is shown to improve with entanglement, but the improvement is with respect to an imperfect experiment with a poor uncertainty. When compared against the SQL, an improvement is not achieved. \label{item_imperfect} \\

\item \textbf{Reduced probability:} The uncertainty from a successful measurement surpasses the SQL, so we have $\Delta \hat{x}_s < \frac{\lambda}{\sqrt{N}T}$, where $\hat{x}_s$ is the estimate post-selected for successful measurements. But the probability to obtain a successful measurement is reduced, such that $\Delta \hat{x} > \frac{\lambda}{\sqrt{N}T}$ considering the total amount of time and resources. \label{item_probability} \\

\item \textbf{Hidden resources:} Extra resources are used to prepare the entangled state, but these are not explicitly accounted for. When accounting for these resources, the uncertainty is worse than the SQL.\label{item_resources} \\

\item \textbf{Noise not uncertainty:} The noise is shown to reduce when entanglement is used. However the resulting uncertainty is not reduced. As the uncertainty comes from the signal-to-noise ratio, simply showing reduced noise does not imply reduced uncertainty, unless one can show that the signal is not reduced. \label{item_noise}\\

\item \textbf{Prior information:} An uncertainty below $\Delta \hat{x} = \frac{\lambda}{\sqrt{N}T}$ is obtained because the prior uncertainty in $\hat{x}$, before the measurement started, is already below the limit. \label{item_prior} \\

\item \textbf{Apples and oranges:} The uncertainty goes below $\Delta \hat{x} = \frac{\lambda}{\sqrt{N}T}$ but for an experiment where this is not the SQL (actually ref.\,\cite{McGuinness2021} is the only example that I know of). This is because the estimated parameter is not in the same class as $x$, and when compared to the correct SQL, the experiment does not exceed this limit. In particular better than $\sqrt{N}$ improvement is not achieved. \label{item_apples} \\

\item \textbf{Substitution and obfuscation:} A quantum sensor is defined containing $N$ resources, then some quantum phase of the entangled sensor is measured and this is compared to a different quantum phase of an unentangled sensor. The resulting uncertainty is then related to some independent physical parameter without ever measuring that parameter. In the end a series of substitutions are performed so that the estimation problem is never clearly defined and different SQLs are conflated with each other.

For certain parameters, such as the polar angle $\theta$ (azimuthal angle $\phi$) of the quantum state in the Bloch sphere, i.e. angle of rotation by a resonant field (quantum phase), then the SQL is often written as $\Delta \hat{x} > \frac{1}{\sqrt{N}}$ and there is no time-dependence. Physically we must require that at some timescale, more time leads to a lower uncertainty. However, for estimation of the quantum phase, the timescale relates to readout of the quantum state and not the coherent evolution time which requires a completely different analysis to what we have considered until now. But there are further issues with this formulation of the SQL. What about performing more than one measurement to further reduce the uncertainty, is this allowed? The answer requires further assumptions on whether the quantum state can be prepared again. There has been a subtle, and unacknowledged substitution in this SQL, where $N$ no longer refers to the number of sensors, but the number of measurements or the number of copies of the parameter that are available. In this respect, the SQL relates to a completely different estimation problem and $N$ defines an entirely different set of resources to the analysis we have performed until now.

However, even if we allow this definition of the SQL and acknowledge that it relates to a different estimation problem where we are only allowed to perform one measurement on each of the $N$ particles, we can also show that entanglement results in a worse uncertainty. First we clarify that the quantum state of an entangled system is an entirely different object to the quantum state of the non-entangled system, thus different parameters are being estimated in each scenario. Let us call $\phi_{\mathrm{NE}}$, the quantum phase of a single atom in the non-entangled system, and $\phi_{\mathrm{E}}$ the quantum phase of the entangled state. Allowing the conventional definition of the SQL, we have $\Delta \hat{\phi}_{\mathrm{NE}} > \frac{1}{\sqrt{N}}$ and $\Delta \hat{\phi}_{\mathrm{E}} > 1$, thus in fact, entanglement provides a worse uncertainty than independent sensors. The common objection is that $\phi_{\mathrm{E}} = N\phi_{\mathrm{NE}}$, and that's how the entangled state obtains an uncertainty of $\Delta \hat{\phi} \approx \frac{1}{N}$. I don't even know how to respond to that statement. What phase is now being measured? Are you saying that an entangled state can measure the quantum phase of some hypothetical non-entangled state better than the same non-entangled state can measure it's own phase? That seems to be the claim. What is clear though, is that the phase of the entangled state is not estimated with a precision $\Delta \hat{\phi}_{\mathrm{E}} < \frac{1}{\sqrt{N}}$, for the uncertainty of a single measurement must be greater than unity.

I expect that what is usually meant, can be expressed in the following proposition. The proposition is that $\phi_{\mathrm{NE}}$ and $\phi_{\mathrm{E}}$ are both related to some classical parameter; for example, rotation through a polar angle $\theta$ depends on the amplitude of a resonant driving field, whereas $\phi$ depends on the phase of the driving field. These are the actual parameters that can be estimated with enhanced precision by exploiting entangled states. However, as we show below and in the literature review, when the uncertainty in these parameters is evaluated, the SQL is not surpassed. \label{item_self} \\
\end{enumerate}

\textbf{Background theory for atoms, ions, spins and atomic clock measurements}

\noindent The review starts with atomic clocks because the number of atoms and experimental time is usually explicitly provided, allowing one to compare to the theoretical limits. We will make use of the following Hamiltonian (c.f. Eq.\,\eqref{eq:Ham}) describing the interaction of a spin with a near-resonant field from time $t_0$ to $t$, as it often occurs in atomic clock literature:
\begin{equation}
H(t_0,t) =\frac{\hbar \omega_0}{2}\sigma_z + \frac{\hbar \Omega}{2} \left( \cos \left[ \omega t + \varphi(t_0) \right] \sigma_x + \sin \left[ \omega t + \varphi(t_0) \right] \sigma_y \right).
\label{eq:clocks}
\end{equation}

\noindent Often $\varphi(t_0)$ is just written as $\varphi$ and the states of the atom measured along $z$ are denoted $\ket{0}, \ket{1}$. If a control field, of known frequency or phase is used to generate Hamiltonian\,\eqref{eq:clocks}, we use the notation $\omega_c$, $\varphi_c$ to imply that the values of these parameters are known and under the experimentalist's control.\\

\textbf{Estimating the laser phase, $\varphi$:}
It is worthwhile to derive the SQL for estimation of the near-resonant laser/microwave phase $\varphi$, as it is incorrectly given in the literature. The SQL for estimating the quantum phase of the atom $\phi$ is \emph{not} necessarily the same as the uncertainty in estimating the laser phase. A common inaccuracy is to state that the uncertainty in estimating $\varphi$ does not depend on the measurement time. Including time in the analysis of $\Delta \hat{\varphi}$ takes a little bit of work. For a single atom, the minimum uncertainty with which one can estimate the phase of the field in the Hamiltonian\,\eqref{eq:clocks} is bounded by:
\begin{equation}
\Delta (|\hat{\varphi}-\varphi_c|)^1 > 1/(\Omega T),
\label{eq:phase}
\end{equation}
where the estimate is with respect to the phase of a control field, and which could be saturated if perfect instantaneous measurements were possible.

Physically, it makes sense that the limit (Eq.\,\eqref{eq:phase}) should depend on the interaction strength of the sensor with the signal, since one would expect that it is easier to estimate the phase of a strong field than a weak field, and a spin with higher gyromagnetic ratio is a better sensor. And of course a longer measurement time should provide a better uncertainty. The limit can be derived following the analysis of refs.\,\cite{Pang2014, Pang2017}. It is a non-constructive method which  provides (a bound for) the maximum quantum fisher information on $\varphi$, $\mathrm{QFI_{max}}(\varphi)$, from the derivative of the Hamiltonian with respect to $\varphi$, and assumes the spin is in a pure state:
\begin{equation}
\text{QFI}_{\max }(\varphi)=\text{  }\left[\int _0^T\text{d$\tau $}\left.\left(\Lambda _{\max }(\tau )-\Lambda _{\min }(\tau )\right)\right/\hbar
\right]^2,
\end{equation}
where $\Lambda _{\max}$ ($\Lambda _{\min }$) are the maximum (minimum) eigenvalues of \(\partial _{\varphi}H(\varphi)\).

An explicit protocol can be found that would saturate the bound of Eq.\,\eqref{eq:phase} assuming perfect measurements could be performed. The method requires that the signal is perfectly resonant to the atom ($\omega = \omega_0$), and requires application of a secondary field also of frequency $\omega_c = \omega_0$ and known phase $\varphi_c$ (all experiments with atoms/ions make use of comparable resources). Simultaneous application of the signal and control field to an atom starting in state $\ket{0}$, followed by a single direct measurement of the atomic population results in a probability to readout the atom in state $\ket{1}$ as a function of the interaction time $t$ given by: $\mathrm{P}(\ket{1}) = \sin\left[\Omega t \cos\left[|\varphi - \varphi_c|/2\right]\right]^2$, where $\Omega$ is the amplitude of the signal and control fields. This probability function saturates the bound when $|\varphi - \varphi_c| = \pi$, and provides information about the relative phase difference between the signal and control fields, and not the difference between the signal phase and the atomic phase. A related method with a factor of $\pi$ worse sensitivity was demonstrated in ref.\,\cite{McGuinness2021}. Rather than estimating an unknown global phase, the bound gives the sensitivity that one can detect a small phase shift from a known phase (again this characteristic is shared by atomic clock experiments).

With $N$ atoms we have:
\begin{equation}
\Delta \hat{\varphi}_{\mathrm{SQL}}^{\mathrm{N}} > \frac{1}{\sqrt{N}\Omega T}.
\label{eq:phaseN}
\end{equation}\\

\textbf{Estimating the laser frequency $\omega$: }
Just as was observed for estimation of $\varphi$, we should also require that $\Delta \hat{\omega}$ depends on the amplitude of the laser and interaction with the sensor. From Eq.\,\eqref{eq:phaseN} and using $\hat{\omega}=\frac{\varphi(t_0)-\hat{\varphi}(T)}{T}$ for known $\varphi(t_0)$ we have:
\begin{equation}
\Delta \hat{\omega}_{\mathrm{SQL}}^{\mathrm{N}} > \frac{1}{\sqrt{N}\Omega T^2}.
\label{eq:freqN}
\end{equation}
However we have been a bit loose with the definition of $\hat{\varphi}(T)$, and the bound can be tightened to \cite{Schmitt2021}: 
\begin{equation}
\Delta \hat{\omega}_{\mathrm{SQL}}^{\mathrm{N}} > \frac{\pi}{\sqrt{N}\Omega T^2},
\label{eq:tightfreqN}
\end{equation}
and further tightened to \cite{McGuinness2021}: 
\begin{equation}
\Delta \hat{\omega}_{\mathrm{SQL}}^{\mathrm{N}} > \frac{2\pi}{\sqrt{N}\Omega T^2}.
\label{eq:tightfreq2N}
\end{equation}
Assuming the 1TF conjecture holds, this scaling cannot continue for a time longer than $T=\pi/\Omega$, i.e. the time to perform a $\pi$-rotation on the atom. For times longer than $T=\pi/\Omega$, Ramsey interferometry provides a linear uncertainty improvement (see Fig.\,\ref{fig2}d).  In this respect, Eq.\,\eqref{eq:tightfreq2N} is not in conflict with the uncertainty limits derived for atomic clocks, since the limits for Ramsey interferometry are derived assuming the $\pi$-rotation time is short compared to the free precession time.

Note, here we are assuming that heterodyning measurements cannot be performed to further reduce the measurement uncertainty below $\frac{1}{\sqrt{N}T}$. In effect we are saying that the $\omega$ is estimated with respect to the atomic larmor frequency $\omega_0$, or we are bounding the measurement uncertainty from a single measurement. For times longer than $\pi/\Omega$, the limit that we will compare to is:

\begin{equation}
\Delta \hat{\omega}_{\mathrm{SQL}}^{\mathrm{N}} > \frac{1}{\sqrt{N} T}.
\end{equation}

\textbf{Estimating the rotation angle, $\theta$ of a quantum state: }
An uncertainty of $\Delta \hat{\theta} \approx \frac{1}{N}$ is often reported for estimating the polar angle of an entangled ensemble. The correct uncertainty relation for a single measurement (even assuming the HL can be saturated) is $\Delta \left(\hat{\theta} - N\Omega T\right) > 1$ (see Loophole \ref{item_self}), where even if $N$, $\Omega$ and $T$ are known perfectly, $\Delta \hat{\theta}$ is greater than unity. Here, we are also assuming only one preparation of the quantum state is possible with $N$ entangled particles. Thus even assuming the HL we can see that entanglement provides a $\sqrt{N}$ worse estimation compared to estimating the angle of $N$ unentangled atoms, the reason being that only one sample can be performed on the entangled ensemble, compared to $N$ independent samples on the unentangled ensemble:
\begin{equation}
\Delta \hat{\theta}_{\mathrm{SQL}}^{\mathrm{N}} > \frac{1}{\sqrt{N}}.
\label{eq:rotN}
\end{equation}

\emph{But, everyone objects, the value of $\theta$ for the entangled state is $N$-fold greater than the unentangled state. This can be used to improve the estimation of the field which drives the rotation angle.} Firstly, this is not the claim that many start with; secondly, the experiments generally do not perform this estimation; thirdly, if that is the claim then the limits for this estimation task should be derived and used as a benchmark.

The SQL for estimating the amplitude of the drive field can be determined by considering magnetic interaction of a single atom with a perfectly resonant field, $\hat{\theta}$ is related to estimating the amplitude of the field, $B_{\perp}$ via the relation $\theta = \Omega T = \gamma B_{\perp} T/\hbar$ (where $\gamma B_{\perp}$ should be replaced with $d E_{\perp}$ for electric dipole interaction). For an atom starting in state $\ket{0}$, measurement of the population after evolution under Eq.\,\eqref{eq:clocks} for time $T$ gives:
\begin{equation}
\Delta \hat{\theta}^1 > 1, \quad \quad \Delta \hat{\Omega}^1 > \frac{1}{T}, \quad \quad \Delta (\hat{B}_{\perp})^1 > \frac{\Omega \hbar}{\gamma T}.
\label{eq:amp}
\end{equation}

Thus we have the SQL for estimating the field amplitude with $N$ atoms:
\begin{equation}
\Delta \hat{\Omega}_{\mathrm{SQL}}^{\mathrm{N}} > \frac{1}{\sqrt{N}T}, \quad \quad \Delta (\hat{B}_{\perp})_{\mathrm{SQL}}^{\mathrm{N}} > \frac{\Omega \hbar}{\sqrt{N}\gamma T}.
\label{eq:ampN}
\end{equation}

\newpage
\subsection*{Experimental Demonstration of Entanglement-Enhanced Rotation Angle Estimation Using Trapped Ions. \emph{Phys. Rev. Lett.} (2001) \href{https://doi.org/10.1103/PhysRevLett.86.5870}{10.1103/PhysRevLett.86.5870}}

Meyer et. al. use two atoms which interact with a near-resonant field of the form of Eq.\,\eqref{eq:clocks}. By entangling the atoms, the atomic rotation angle $\theta$, the phase of the field $\varphi$, and the frequency difference $|\omega - \omega_0|$ is estimated. The claim presented by Meyer et. al. is that an uncertainty below the standard quantum limit in estimating these parameters is obtained when the two atoms are entangled.\\

\textbf{Loophole \ref{item_time}:} Time is not included in the analysis. An uncertainty in estimating the phase $\varphi$, of the laser is shown to go below the limit: $\Delta \hat{\varphi} < 1/\sqrt{N}$. For estimating the frequency difference between the laser and the atomic Larmor an uncertainty of $\Delta (\omega - \omega_0) < 1/(\sqrt{N}t)$ is also demonstrated, where only the phase evolution time is considered, not the entire measurement time. When the entire measurement time is included in the analysis, the uncertainty does not go below the SQL.\\

\noindent \emph{Estimation of the laser phase $\varphi$}

The actual SQL for estimation of the laser phase, which cannot be surpassed with measurements on $N$ independent particles is: $\Delta \hat{\varphi}_{\mathrm{SQL}}^{\mathrm{N}} > \frac{1}{\sqrt{N}\Omega T}$ (Eq.\,\eqref{eq:phaseN}). This uncertainty can go below $1/\sqrt{N}$ when $\Omega T > 1$ and does require entanglement.

\textbf{\textcolor{NavyBlue}{Analysis:}} Measurement of the laser phase requires knowledge of the Rabi frequency which is not explicitly given in the paper. Meyer et. al. cite refs. [16, 18] of their paper for experimental details of the driving fields. In these references a $\pi$-rotation time of $\sim 2.5\,\mu$s is achieved, giving $\Omega \sim 1.3$\,MHz. For the reported experimental duration of $\simeq 1$\,ms and for $N = 2$, the SQL is $\Delta \hat{\varphi}_{\mathrm{SQL}}^2 > 1/(\sqrt{2} \times (1.3 \times 10^6) \times 0.01) \simeq 5\times 10^{-4}$. This limit is far below the demonstrated uncertainty with entanglement: $\Delta \hat{\varphi} = 0.59$.

\noindent\textcolor{Maroon}{Uncertainty below $N = 1$ limit? No.}\\

\noindent \emph{Estimation of the laser frequency $\omega$}

Meyer et. al. also estimate the frequency difference between the laser and atomic resonance, showing $\Delta (\hat{\omega} - \omega_0) < 1/(\sqrt{N}t)$ where $t$ is the phase evolution time.

\textbf{\textcolor{NavyBlue}{Analysis:}} Specifically, an uncertainty of $\Delta (\hat{\omega} - \omega_0) \approx 5.5 \times 10^4$\,Hz is achieved for a phase evolution time $t \approx 11.5\,\mu$s. Considering the total experimental duration $\simeq 1$\,ms, the uncertainty is worse than the SQL for two atoms: $\Delta (\hat{\omega} - \omega_0)_{\mathrm{SQL}}^2 > 707$\,Hz.

\noindent\textcolor{Maroon}{Uncertainty below $N = 1$ limit? No.}\\

\noindent \emph{Estimation of atomic rotation angle $\theta$ and Rabi frequency $\Omega$}

\textbf{Loophole \ref{item_self}:} An uncertainty in estimating the angle $\theta$, of atomic rotation from the $z$-axis, is reported to go below the limit: $\Delta \hat{\theta} < 1/\sqrt{N}$.

\textbf{\textcolor{NavyBlue}{Analysis:}} We disagree, in fact Meyer et. al. demonstrate a measurement uncertainty of $\Delta \hat{\theta} = 2\times 0.65$ with two entangled atoms, taking $\simeq 1$\,ms. This uncertainty is greater than unity and not less than $1/\sqrt{2}$. Accounting for the two-fold enhanced rotation speed does this allow for improved estimation of the field amplitude? Interaction with the field is only for a short duration $\simeq 1.25\,\mu$s, to perform a $\pi/2$-rotation. Therefore the Rabi frequency cannot be estimated with an uncertainty better than: $\Delta \hat{\Omega} > 4 \times 10^5$\,Hz, assuming Heisenberg limited estimation. Considering the total measurement time, the SQL for estimating the Rabi frequency with $N = 2$ is: $\Delta \hat{\Omega}_{\mathrm{SQL}}^2 > 1/(\sqrt{2} \times 0.001) = 707$\,Hz, see Eq.\,\ref{eq:ampN}. The SQL gives an uncertainty limit nearly three orders of magnitude below what could possibly have been experimentally demonstrated with the entangling scheme used by Meyer et. al.

\noindent\textcolor{Maroon}{Uncertainty below $N = 1$ limit? No.}

\newpage

\subsection*{Toward Heisenberg-Limited Spectroscopy with Multiparticle Entangled States. \emph{Science} (2004) \href{https://doi.org/10.1126/science.1097576}{10.1126/science.1097576}}

Leibfried et. al. demonstrate phase estimation with three ions where measurement of the quantum phase of the ions $\phi$, is conflated with measurement of the laser phase $\varphi$. The authors apply the following SQL to estimation of both the laser phase and the quantum phase: $\Delta \hat{\phi} < 1/\sqrt{N}$, where there is no time dependence and it is implicitly assumed that only one measurement can be performed on each spin. Why can't, for example, more than one measurement be performed on each spin, allowing better than $1/\sqrt{N}$ uncertainty? Not only that, an uncertainty much less than $1/\sqrt{N}$ can also be obtained from a single measurement on $N$ independent spins, if the laser phase is known to be within some small initial range. The following uncertainty limit in estimating the laser phase, however, cannot be surpassed with $N$ independent spins: $\Delta \hat{\varphi}_{\mathrm{SQL}}^{\mathrm{N}} > \frac{1}{\sqrt{N}\Omega T}$ (Eq.\,\eqref{eq:phaseN}).

By relating phase estimation to frequency estimation via $\Delta \hat{\varphi} = \Delta (\hat{\omega} - \omega_0)t$, Leibfried et. al. also connect improvements in phase estimation to improvements in frequency estimation. Ultimately, by entangling the three ions, the authors claim to achieve a spectroscopic sensitivity enhanced by a factor of 1.45 compared to a perfect experiment with non-entangled ions. Below we show that Leibfried et. al. do not surpass the SQL, and do not outperform the sensitivity of a perfect experiment with non-entangled ions. At least for any normal understanding of that sentence.\\

\noindent \emph{Estimation of the laser phase $\varphi$}

\textbf{Loophole \ref{item_time}:} Leibfried et. al. measure the phase $\varphi_2$, of a second $\pi/2$-pulse with respect to the phase $\varphi_1$, of an initial $\pi/2$-pulse, from which a phase sensitivity a factor of $0.84 \times \sqrt{3}$ higher than the SQL is claimed. This would mean that they achieved a phase uncertainty: $\Delta \hat{\varphi} = 1/(3 \times 0.84) = 0.4$. Sometimes $\varphi_1,\varphi_2$ refer to the phase of the entangling, disentangling pulses, but that is inconsequential.

\textbf{\textcolor{NavyBlue}{Analysis:}} Leibfried et. al. clearly provide all the experimental parameters allowing one to perform a comparison to the actual SQL (Eq.\,\ref{eq:phaseN}). A pulse sequence taking 43\,$\mu$s performs a $2\pi$-rotation and includes two $14\,\mu$s delays (one for a phase gate). Therefore the Rabi frequency of the laser is $\Omega \approx 400$\,kHz. The total experimental duration was $(43 + 2 + 43 + 400)\,\mu\mathrm{s} = 488\,\mu$s, where we have assumed the atoms can be instantaneously initialised into the state $\ket{\downarrow, N}$ via optical pumping, since that timescale was not provided in the manuscript. The SQL is $\Delta \hat{\varphi}_{\mathrm{SQL}}^3 > \frac{1}{\sqrt{3} (488 \times 10^{-6}) (400 \times 10^3)} = 0.003$. How about if we also neglect the 400\,$\mu$s taken to readout the ions? Then the SQL is $\Delta \hat{\varphi}_{\mathrm{SQL}}^3 > \frac{1}{\sqrt{3} (88 \times 10^{-6}) (400 \times 10^3)} = 0.02$, again much less than the uncertainty Leibfried et. al. demonstrated.

While we agree that if much more time is allocated to the experiment using entanglement, then Leibfried et. al. demonstrated a phase sensitivity that cannot be achieved with independent ions. This is the only way to restrict a perfect measurement with independent ions from achieving a better uncertainty. But if that is the case, what do we need entanglement for? By allocating more time to one experiment with independent ions, one can also achieve a better uncertainty than the same experiment in less time.\\
\noindent\textcolor{Maroon}{Uncertainty below $N = 1$ limit? No.}\\

\noindent \emph{Estimation of the laser frequency $\omega$}

\textbf{Loophole \ref{item_time}:} From the claimed improvement in phase sensitivity, Leibfried et. al. imply that (in principle) an improved frequency sensitivity can be obtained. However, no time analysis is included in the presentation of these improvements.

\textbf{\textcolor{NavyBlue}{Analysis:}} We can also bound the uncertainty in frequency estimation that Leibfried et. al. can obtain. The time between the two $\pi/2$-pulses is 6\,$\mu$s: given by the time to perform a $\pi/2$-pulse plus an additional 2\,$\mu$s wait time between pulses. Therefore the frequency uncertainty cannot be less than $\Delta (\hat{\omega}-\omega_0) > \frac{\Delta\hat{\varphi}}{t} = \frac{0.4}{6 \times 10^{-6}}\,\mathrm{Hz} = 70$\,kHz. As the total experimental duration (excluding initialisation) was $488\,\mu$s, the SQL is $\Delta (\hat{\omega}-\omega_0)_{\mathrm{SQL}}^3 > \frac{1}{\sqrt{3}\times 488 \times 10^{-6}}\,\mathrm{Hz} = 1$\,kHz. If we would also neglect the time to readout the ions, we have that the SQL is $\Delta (\hat{\omega}-\omega_0)_{\mathrm{SQL}}^3 > \frac{1}{\sqrt{3}\times 88 \times 10^{-6}}\,\mathrm{Hz} = 7$\,kHz.\\
\noindent\textcolor{Maroon}{Uncertainty below $N = 1$ limit? No.}\\

\noindent \emph{Estimation of the atomic phase $\phi$}

\textbf{Loophole \ref{item_self}:} Leibfried et. al. might also claim that irregardless of the measurement time, the enhanced laser phase estimation \emph{per measurement}, provides an uncertainty in estimating the quantum phase of the sensor which is below the SQL: $\Delta \hat{\phi} > 1/\sqrt{N}$.

\textbf{\textcolor{NavyBlue}{Analysis:}} Assuming that the laser phase is known, and the quantum phase is estimated relative to this parameter, we have the uncertainty in estimating the quantum phase of an entangled system with a single measurement is bounded by: $\Delta \left(\hat{\phi} - N\varphi\right) > 1 \quad \rightarrow \quad \Delta \hat{\phi} > 1$. Thus the uncertainty in estimating the phase of the entangled system from a single measurement is bounded by unity. As $N$ independent measurements on the unentangled system can be performed with the same resources, it is clear that entanglement produces a worse uncertainty. Although it is correct that the value of the phase is larger in the entangled system, it is incorrect to say that it can be estimated with enhanced precision compared to $N$ independent particles. In fact, the precision cannot be better than obtained with a single particle.\\
\noindent\textcolor{Maroon}{Uncertainty below $N = 1$ limit? No.}\\

\newpage

\subsection*{Nonlinear atom interferometer surpasses classical precision limit. \emph{Nature} (2010) \href{https://doi.org/10.1038/nature08919}{10.1038/nature08919}}
Gross et. al. perform Ramsey interferometry on several hundred entangled atoms. The idea appears simple, the Ramsey interferometer measures the relative phase difference between the atomic phase $\phi$ of the atoms and the phase $\varphi$ of the $\pi/2$-pulses (note the different notation in the paper where $\varphi$ is the quantum phase). The analysis also appears simple, Gross et. al. say that the uncertainty in this estimate is bounded by the SQL: $\Delta \hat{\phi} > 1/\sqrt{N}$. By entangling the atoms in the interferometer, in particular by generating a coherent spin squeezed state, Gross et. al. claim to achieve an uncertainty of $\Delta \hat{\phi} < 1/\sqrt{N}$. It is not clear if this refers to the phase of the laser or the quantum phase. If the claim is estimation of the quantum phase, then this claim is incorrect as the uncertainty does not go below unity (see Loophole \ref{item_self}). The authors write that the relative phase is proportional to the quantity to be measured (caption Fig. 1) but this quantity is not explicitly defined. We consider both estimation of the phase of the second $\pi/2$-pulse with respect to the known atomic phase, and alternatively estimation of the atomic phase with respect to known laser phase. We show that in both instances Gross et. al. do not demonstrate an uncertainty below the SQL.\\

\noindent \emph{Estimation of the laser phase $\varphi$}

\textbf{Loophole \ref{item_time}:} No time analysis is presented in estimation of the laser phase. We have that the uncertainty in estimating the phase $\varphi$ of the second $\pi/2$-pulse from a single measurement of the entangled state is bounded by $\Delta \left(N\hat{\varphi} - \phi\right) > 1$. If the atomic phase is known from the phase of the first $\pi/2$-pulse, then we obtain the following bound assuming Heisenberg limited phase accumulation: $\Delta \hat{\varphi} > 1/N$. While it may be possible to achieve an uncertainty $\Delta \hat{\varphi} < 1/\sqrt{N}$ from a single measurement, when the time required to achieve this uncertainty is taken into account, the SQL: $\Delta \hat{\varphi} < 1/(\sqrt{N} T)$ is not surpassed. In fact, Gross et. al. do not achieve an uncertainty $\Delta \hat{\varphi} < 1/\sqrt{N}$ even when excluding the measurement time. 

\textbf{\textcolor{NavyBlue}{Analysis:}} Gross et. al. report the following experimental values for estimation of the relative phase of the second $\pi/2$-pulse. $N = 2300/6 = 383$,  $\Omega = 2 \pi \times 600$\,Hz and a total time of 20\,ms. The measurement time is composed of the non-linear beamsplitter duration of 18.3\,ms (Fig.\,2a) an additional $\pi/2$-pulse and a readout duration of approximately 1\,ms. From Eq.\,\ref{eq:phaseN} we have the SQL for estimating the laser phase is: $\Delta \hat{\varphi}_{\mathrm{SQL}}^{383} > \frac{1}{\sqrt{383}\times 2 \pi \times 600 \times 0.02} = 6.8\times10^{-4}$.

In Fig.\,2c, Gross et. al. report a phase uncertainty $\Delta \hat{\varphi} \approx 11$, which is 5 orders of magnitude worse than the SQL. This uncertainty is not less than $1/\sqrt{N}$ even excluding the measurement time, so I don't see how an uncertainty below the SQL can be claimed, or in the words of Gross et. al. how one can claim to achieve a performance ``$15\%$ superior to that in the ideal classical case". An uncertainty better than this value could be obtained with a single atom in the same measurement time.

\noindent\textcolor{Maroon}{Uncertainty below $N = 1$ limit? No.}\\

\noindent \emph{Estimation of the atomic phase $\phi$}

\textbf{Loophole \ref{item_self}:} Gross et. al. also conflate estimation of the laser phase with estimation of the atomic phase.

\textbf{\textcolor{NavyBlue}{Analysis:}} The bound $\Delta \left(N\hat{\varphi} - \phi\right) > 1$ also shows that estimation of the atomic phase is worse with the entangled state, since for a single measurement of the entangled state we have an uncertainty bound $\Delta \hat{\phi} > 1$. With $N$ independent measurements on an unentangled ensemble a better uncertainty can be obtained from the same amount of resources, and indeed an equivalent uncertainty could be achieved with a single particle.

\noindent\textcolor{Maroon}{Uncertainty below $N = 1$ limit? No.}\\

\newpage

\subsection*{Orientation-Dependent Entanglement Lifetime in a Squeezed Atomic Clock. \emph{Phys. Rev. Lett.} (2010) \href{https://doi.org/10.1103/PhysRevLett.104.250801}{10.1103/PhysRevLett.104.250801}}

Leroux et. al. squeeze the state of $3.5 \times 10^4$ effective atoms (I am not sure what effective means) and perform Ramsey interferometry on these atoms to determine the Allen deviation in the frequency estimation of a near-resonant field. Their experimental uncertainty is compared to a different standard quantum limit which depends on square-root of the measurement time. In particular the SQL that Leroux et. al. refer to is:
\begin{equation}
\Delta \left(\hat{\omega}-\omega_0 \right) = \frac{1}{t}\sqrt{\frac{T_c}{N T}} \; \mathrm{Hz}/\sqrt{T},
\label{eq:leroux}
\end{equation}
where $t (T_R)$ is the time in the Ramsey interferometer, $T (\tau)$ is the total measurement time, $T_c$ is the clock cycle time which includes clock readout and preparation times, $N (2 S_0)$ the number of atoms, and the parenthesis indicate the different labelling used by Leroux et. al. in their paper. Note, we have removed the factor of $1/\omega_0$ in Eq.\,\ref{eq:leroux} so that the absolute frequency uncertainty is expressed rather than the fractional frequency uncertainty. While we agree that, for the particular configuration of atomic clock Leroux et. al. employed, including all imperfections and a $200\,\mu$s coherence time, this equation sets the achievable experimental uncertainty, we disagree that this is the standard quantum limit. In particular, we disagree that this equation limits the ultimate precision achievable with the same $N$ unentangled atoms and measurement time $T$. For example, the duty cycle of Leroux et. al.'s clock was $2 \times 10^{-5}$, and improvement in this parameter would provide several orders of magnitude enhanced precision without requiring entanglement. In the following analysis we show that Leroux et. al. do not demonstrate a frequency uncertainty below the SQL (Eq.\,\ref{eq:SQL}) when entangling $N$ atoms, or even below that achievable with a single atom.\\

\textbf{Loophole \ref{item_time}, \ref{item_multiple}:} Time is only partially included in the comparison to the SQL. The formula Leroux et. al. present for the SQL has square-root dependence on the measurement time (see page 3 in their paper where measurement time is $\tau$). This formula requires that multiple independent measurements are performed during the measurement time. Specifically, Leroux et. al. perform Ramsey measurements on $3.5 \times 10^4$ effective atoms, where each measurement sequence takes 9 seconds and the phase evolution time in the Ramsey interferometer is $200\,\mu$s. They report an experimental uncertainty of:
\begin{equation}
\Delta \left(\hat{\omega}-\omega_0 \right) = 12.6/\sqrt{T} \; \mathrm{Hz}, 
\end{equation}
for measurement times in the range of 9 - 100 seconds\footnote{Thank-you Leroux et. al. for presenting all of the data and analysis so clearly!!! Note, I think I have the factor of $2 \pi$ correct.}.

\textbf{\textcolor{NavyBlue}{Analysis:}} For $N = 3.5 \times 10^4$, this uncertainty is not less than $\Delta (\hat{\omega}-\omega_0)_{\mathrm{SQL}}^N > \frac{1}{\sqrt{3.5 \times 10^4} T}$, over the experimental range $9 \leq T \leq 100$ seconds. In fact this uncertainty is not less than $1/T$ for any point in the experimental range.

\noindent\textcolor{Maroon}{Uncertainty below $N = 1$ limit? No.}

\newpage

\subsection*{Quantum spin dynamics and entanglement generation with hundreds of trapped ions. \emph{Science} (2016) \href{https://doi.org/10.1126/science.aad9958}{10.1126/science.aad9958}}

I don't know how to critique this paper, it is so opaque to me that it borders on impenetrable. Bohnet et. al. claim 4\,dB of spectroscopic enhancement. Spectroscopy of what? A reduced spin variance as a function of tomography angle $\psi$ for $N = 82$ atoms is claimed. Um... ok, I am not sure how that demonstrates spectroscopic enhancement. I am being deliberately obtuse here. It is clear that the authors demonstrate entanglement between large numbers of atoms, they can quantify this entanglement and observe non-classical statistics. I can't argue with that and it sounds impressive. It is also not my area of expertise and that might explain why I have no idea what they are talking about. But quantum sensing is an area I specialise in, and I have no idea how an enhanced sensing precision can be claimed. This is perhaps not the central claim of the manuscript, but it is also not a minor point and significant space is allocated to discussion of enhanced spectroscopic precision through entanglement.\\

\textbf{Loophole \ref{item_noise}:} I don't know how to categorize this one. Showing a measurement of reduced spin variance doesn't mean anything in terms of metrology. Just put all your spins in the state $\ket{0}$ and then measure them. The variance of this measurement is much less than $\sqrt{N}$, but have you measured anything interesting. I don't think so.

\textbf{\textcolor{NavyBlue}{Analysis:}} Maybe the spin state that Bohnet et. al. generate is also supposed to have an enhanced sensitivity to some parameter like the tomography angle, or the angle $\theta$ (see Fig. 4C), but I couldn't find anywhere in the paper where $\theta$ is defined, so I don't know what this is. It is also not clear to me, but if I understand correctly, the experimental value of the Fisher information divided by $N$ does not exceed the classical limit ($F/N = 0.56$). So putting aside my objection that the Fisher information is time-independent, no experimental enhancement is observed.

\noindent\textcolor{Maroon}{Uncertainty below $N = 1$ limit? No idea.}
\newpage




\subsection*{Toward Heisenberg-Limited Rabi Spectroscopy. \emph{Phys. Rev. Lett.}, (2018) \href{https://doi.org/10.1103/PhysRevLett.120.243603}{10.1103/PhysRevLett.120.243603}}

Shaniv et. al. experimentally estimate the average transition frequency of two ions, $\omega_0 = (\omega_0^1 + \omega_0^2)/2$, where $\omega_0^1,\; \omega_0^2$ denote the transition frequency of the two ions (I think this is what is meant by average transition frequency, even though it is defined $\omega_0 = (\omega_0^1 + \omega_0^2/2)$ in their paper). This is related to the detuning from the laser frequency: $\delta_1 = \omega -\omega_0$. A similar analysis is done for estimating half the frequency difference: $\delta_2 = (\omega_0^1 + \omega_0^2)/2$, where again, I am assuming they have a typo in their definition. By entangling the ions, Shaniv et. al. report a frequency spectrum, where the linewidth of the population vs. $\delta_1$ is approximately half that obtained when the ions are not entangled, likewise for the $\delta_2$ spectrum. From analysis of the spectra, a frequency uncertainty below the SQL is claimed.\\

\textbf{Loophole \ref{item_time}, \ref{item_multiple}, \ref{item_imperfect}:} By analysing the data presented in Fig. 4, Shaniv et. al. report a frequency sensitivity of $0.23\,\mathrm{Hz}/\sqrt{\mathrm{Hz}}$ when entangling the ions, compared to a sensitivity of $0.43\,\mathrm{Hz}/\sqrt{\mathrm{Hz}}$ for independent ions. The claim is that these results are ``well below the below the standard quantum limit and close to the Heisenberg limit". First, showing an improvement in sensitivity is not proof you have achieved a sensitivity below the SQL, unless you improved upon a sensitivity already at the SQL (Loophole \ref{item_imperfect}). Also, how can one achieve better than $\sqrt{2}$ improvement with two atoms if the sensitivity was already at the SQL? Not only that, the quoted sensitivity would outperform not only state-of-the-art atomic clocks but also the Heisenberg limit for experimental times up to several seconds, so the numbers seem wildly incorrect.

Second, Shaniv et. al. do not explicitly define the SQL in their paper, so it is not clear what limit they claim to outperform, however the units would indicate an uncertainty of: $\omega > \frac{\alpha}{\sqrt{N T}}$, for some parameter $\alpha$. The only sensible analysis of their results, is that the sensitivity is expressed solely in terms of the interaction time, and does not include any other readout or preparation overheads. Furthermore Shaniv et. al. assume multiple independent measurements are being performed leading to $\sqrt{T}$ improvement. It may be true that an uncertainty below this limit is experimentally achieved, however we disagree that such an uncertainty cannot also be surpassed without entangling the atoms as long as all overheads are explicitly included. The correct limit should be Eq.\,\eqref{eq:SQL}, and we assert that this cannot have been exceeded in the experiment.

\textbf{\textcolor{NavyBlue}{Analysis:}} The experimental times required to obtain the reported frequency sensitivities are not provided in the paper, thus we are not able to conduct an analysis.

\textbf{Author response:}\emph{Through private communications, the authors confirm that the published uncertainties are incorrect.}

\noindent\textcolor{Maroon}{Uncertainty below $N = 1$ limit? The paper reports an uncertainty beyond this limit, but the Author response confirms that the data is incorrect.}\\

\newpage

\subsection*{Optimal frequency measurements with quantum probes. \emph{npj Quantum Information} (2021) \href{https://doi.org/10.1038/s41534-021-00391-5}{10.1038/s41534-021-00391-5}}

Since I have gone to great effort in publicly critising other authors for making claims that do not stack up, it is only fair that I critique a paper that I have authored. You know, don't throw stones, and all that. In Schmitt et. al., on which I act as corresponding author, we claim to demonstrate frequency discrimination better than $1/T$ and also a frequency uncertainty that improves quadratically with measurement time. Although it is not explicitly claimed in the paper, the impression is that this frequency uncertainty can continue indefinitely and therefore one would significantly outperform the Heisenberg limit when using the demonstrated techniques (see Eq.\,8 of Schmitt et. al.). In fact there is simply no discussion of how long this scaling can continue, the possibility was not addressed in the paper. \\

\emph{Estimation of the microwave frequency $\omega$}

\textbf{Loophole \ref{item_time}, \ref{item_scaling}:} With a single spin, Schmitt et. al. demonstrate a frequency uncertainty of: $\Delta \omega = \frac{\alpha}{t^2}$ (see Fig.\,6\,e,f)), where only the phase evolution time is considered and not the total experimental duration.

\textbf{\textcolor{NavyBlue}{Analysis:}} For the experimental data presented $\alpha > t$, so although the frequency uncertainty shows an enhanced scaling, it never goes below $1/T$. Actually $\alpha$ is explicitly defined as $\pi/\Omega$ (Eq.\,(8) of the paper, note factor of two difference in definition of $\Omega$ used in Schmitt et. al. compared to the definition used in this manuscript), and as the interaction time is less than the time to perform a $\pi$-rotation we have $\alpha/t > 1$.

\noindent\textcolor{Maroon}{Uncertainty below $N = 1$ limit? There was just one atom, so no.}\\

\emph{Discrimination between two frequencies, $\omega_1$ and $\omega_2$}

\textbf{Loophole \ref{item_prior}:} Schmitt et. al. also demonstrate frequency discrimination between two frequencies, where the frequency difference between the frequencies is a factor of 10 than the inverse of the interaction time (not the entire measurement time).

\textbf{\textcolor{NavyBlue}{Analysis:}} Discrimination relies on prior knowledge, and thus the uncertainty reduction from this measurement does not break the SQL. In fact discrimination allows an uncertainty reduction, from an initial prior uncertainty, by a factor of two at most (see \emph{Prior Information} section of \ref{App:SQL}). In addition, only the phase evolution time was considered in the analysis, when taking into account the readout time, the SQL is not exceeded.\\ 

Author response: \emph{I believe at the time of writing that we assumed it would continue indefinitely, and did not consider that this may not be the case. Only later, have I begun to question this assumption and looked into experimental confirmation of this assumption. Although the tone of these reviews are antagonistic, this is not intended. I am trying to be critical and point out errors or misleading statements in all reviewed papers. I am aware that it is actually extremely difficult to publish a paper without any errors or misleading statements - my current most cited paper could be argued to have an incorrect or misleading title. My most recent publication also has an incorrect title. I do not consider myself to upheld a higher scientific standard in my work, than the reviewed papers.}

\newpage


\textbf{Literature on photon measurements}

The following papers in the literature review concern quantum measurements with photons. This is not my area of expertise, but then again, neither are atomic clocks. Two areas that I have found particular difficulty in performing the analysis are: \emph{i)} defining the experimental time, and \emph{ii)} determining the correct parameter $\lambda$ to define the SQL (see Eq.\,\eqref{eq:SQL}). Another issue is that the photon number in the experiment is often not as well characterised (at least not as well as in the atomic clock literature).

The experimental time should depend on the time to generate the photons, the time that the photons spend in the interferometer (proportional to the length) and the time to detect the photons (dependent on the detector bandwidth). In general these parameters are not provided for squeezed experiments, in particular, the preparation overheads are generally excluded. We argue, just as for Ramsey interferometry with atoms, that with photon interferometry we need to consider more than just the input state into the interferometer, we also need to consider the time and resources to prepare the input state and the time detect the output state.

\newpage

\subsection*{Precision measurement beyond the shot-noise limit. \emph{Phys. Rev. Lett.} (1987) \href{https://doi.org/10.1103/PhysRevLett.59.278}{10.1103/PhysRevLett.59.278}}

Xiao et. al. perform Mach-Zehnder interferometry by inserting a coherent field into one of the input ports of the interferometer (and a vacuum into the other). The number of photons in this field has fluctuations on the order of $\sqrt{N}$, where $N$ is the mean number of photons per second (Xiao et. al. write $P$ instead). The phase of the photons in one of the interferometer arms is modulated by applying a time-dependent voltage to a phase modulator. By measuring the output photon intensity from the interferometer, the phase shift is estimated. The authors assert that the shot-noise limit is the best sensitivity possible for inputs of a coherent state and a vacuum field. The authors also state that by using entanglement in the form of squeezed light, a sensitivity beyond this shot-noise limit is achieved.\\

\textbf{Loophole \ref{item_multiple}, \ref{item_imperfect}:} The shot-noise limit that Xiao et. al. refer to is indeed the best sensitivity possible when accounting for the detector bandwidth, photon losses and imperfect detection efficiency. However, this is not the best sensitivity possible for inputs of a coherent state and a vacuum field. The authors state that the amplitude of the phase shift $\delta$ cannot be estimated better than $\Delta \hat{\delta} = 1/\sqrt{T_B \alpha \xi N} = 1/\sqrt{N_\mathrm{e}}$, where $T_{B}$ is the inverse of the detector bandwidth, $\alpha$ is the quantum efficiency of the photodetector, $\xi$ is an efficiency factor that accounts for possible losses in propagation and $N_\mathrm{e}$ is the mean number of detected photoelectrons. This is an imperfect limit and assumes that multiple independent measurements are performed, limited by the detector bandwidth. A setup with lower photon losses and better detection efficiency would achieve a lower sensitivity than Xiao et. al. achieve (also dictated by the same equation) from the same coherent state and vacuum field without requiring entanglement. Furthermore a setup with longer cavity length and a different detector bandwidth would achieve linear improvement with measurement time from the same coherent state and vacuum field without requiring entanglement, and this sensitivity would outperform the sensitivity that Xiao et. al. achieve with entanglement.\\

\textbf{\textcolor{NavyBlue}{Analysis:}}
For the data shown in Fig.\,2, where Xiao et. al. report a signal enhancement of 3\,dB using squeezed light, the experimental parameters were $\alpha = 0.89$, $\xi = 0.94$.
In addition, the optical power used is 800\,$\mu$W, with a wavelength of 1\,$\mu$m, which gives $5 \times 10^{14}$ photons/sec. The detection bandwidth is 100\,kHz, so each detection takes 10\,$\mu$s, during which $5 \times 10^9$ photons are produced. The arm length of the interferometer was not given, but we assume that the phase modulators completely filled the arms, which gives an arm length of 5\,cm, since this is the approximate length of the crystals. With a refractive index of 1.5, this gives a time of $2.5\times 10^{-10}$ seconds in the interferometer. Therefore, if we would extend the interaction time up to the detector bandwidth, this would provide a 4,000 fold signal enhancement. A further improvement in detection efficiency and losses would give a 4,400-fold or over 36\,dB improvement on the experimental results.

\textbf{Loophole \ref{item_resources}:} A further loophole which is not explicitly discussed is the resources required to generate the squeezed light. In particular, Xiao et. al. do not report the experimental efficiency of the optical parametric oscillator (OPO) which was used to generate the squeezed light. This efficiency is vital to comparisons of the total resources used in the experiment, since those input OPO photons could alternatively have been sent into the interferometer to improve the sensitivity. Therefore we argue that a complete analysis needs to take into account the efficiency of the squeezed light generation. In order to compare to the uncertainty achievable with a single photon, we also need to know, in addition to the efficiency of the OPO, how efficient the squeezing was, and the resulting number of independent photon states in the interferometer at any given time. It is this number of independent photon states that we should compare to.

\noindent\textcolor{Maroon}{Uncertainty below $N = 1$ limit? I am not able to perform this analysis. In particular we need to know the OPO efficiency for generating the squeezed light, and in addition, for squeezed light input into the interferometer, how many independent photon states are in the interferometer at any time, this should define $N$ in the SQL.} \\

A similar analysis can be conducted to show that the following article does not achieve an uncertainty below the SQL.

\subsection*{Spectroscopy with squeezed light. \emph{Phys. Rev. Lett.} (1992) \href{https://doi.org/10.1103/PhysRevLett.68.3020}{10.1103/PhysRevLett.68.3020}}

%


\newpage

\subsection*{Beating the Standard Quantum Limit with Four-Entangled Photons. \emph{Science} (2007) \href{https://doi.org/10.1126/science.1138007}{10.1126/science.1138007}}

Nagata et. al. describe using $N$ particles (photons) as input into an interferometer in order to measure a phase shift $\phi$, which is generated in one arm of the interferometer. They state that the uncertainty with which one can estimate $\phi$ with independent measurements is: $\Delta \hat{\phi} = 1/\sqrt{N}$, whereas by using an entangled $N$-photon state, a superior uncertainty of  $\Delta \hat{\phi} = 1/N$ can be achieved. A minor quibble is that the equality symbols should be replaced with inequality symbols in these uncertainty bounds. A more significant issue with this analysis is that there is no time dependence for the estimation problem, and (in the introduction at least) it is not clear exactly what parameter Nagata et. al. claim to estimate with superior precision -- is it the phase of the quantum state, or is it some parameter of the phase shifter in the interferometer which generates that phase shift? In Figs. 3, 4 Nagata et. al. report on measurements of the angle of the phase-shifter in one arm of the interferometer. Here, we perform the uncertainty analysis for both of these estimation problems and show that Nagata et. al. do not outperform the SQL in either case.\\

\noindent \emph{Estimation of the phase shift imparted onto the quantum state in the interferometer}

\textbf{Loophole \ref{item_self}:} The phase shift in the interferometer sends the state after the first beam-splitter to the state before the second beam-splitter according to: $\left(\ket{N0}+\ket{0N}\right)/\sqrt{2} \rightarrow \left(\ket{N0}+e^{i\phi_N}\ket{0N}\right)/\sqrt{2}$, where $N$ is the number of photons in the (entangled) input state. We can show that the uncertainty in estimating $\phi_N$ does not go below the SQL, even with an entangled input state.

\textbf{\textcolor{NavyBlue}{Analysis:}} For a single measurement on a single quantum system, the uncertainty in estimating the phase of the quantum state is bounded by: $\Delta \hat{\phi}_N > 1$. As only one copy of the $N$-photon entangled state can be generated with $N$ photons, the uncertainty in estimating the phase of this state cannot go below unity. However, $N$ independent copies of the single photon state can be measured, allowing an improved uncertainty to be achieved, bounded by $\Delta \hat{\phi}_N > 1/\sqrt{N}$. While it is true that the magnitude of the quantum phase imparted onto the $N$-photon entangled state is greater than the phase imparted onto the single photon state, it is not correct to say that the uncertainty in estimating that phase is $N$-fold greater when entanglement is used, because it is in fact $\sqrt{N}$ worse.

\noindent\textcolor{Maroon}{Uncertainty below $N = 1$ limit? The uncertainty is bounded by unity, this cannot beat the $N = 1$ limit.}\\

\noindent \emph{Estimation of angle of the phase-shifter in the interferometer}

\textbf{Loophole \ref{item_probability}, \ref{item_resources}:} Taking into account all of the $N$ photon resources that Nagata et. al. have at their disposal, the phase shift in the interferometer sends the state after the first beam-splitter to the state before the second beam-splitter according to: $\left(\ket{I0}+\ket{0I}\right)/\sqrt{2} \rightarrow \left(\ket{I0}+e^{i I x t}\ket{0I}\right)/\sqrt{2}$, where $I$ is the number of photons in the entangled input state which is less than $N$, and now we are explicitly measuring some parameter $x$ of the phase-shifter which interacts with the photons for a duration of $t$. We have that $x$ produces a relative quantum phase on the photon state according to $\phi_I = I x t$, and now the expectation (or imputation by Nagata et. al.) is that this parameter can be estimated with improved precision when exploiting entanglement. However, to achieve an entanglement advantage, the factor of $I$ enhancement provided by the entangled state must compensate for the loss of photons in the generation of the entangled state. Nagata et. al. do not address this question.

In Figs. 3, 4 Nagata et. al. explicitly discuss and report measuring the angle of the phase shifter with enhanced precision, so we shall conduct the analysis where $A \equiv x t$, is the angle of the phase shifter. For simplicity, we exclude increasing the interaction time with the phase shifter as a method to improve the precision, rather, we shall show that the probability to produce and detect the entangled photons results in a worse sensitivity than if the input light to the spontaneous parametric down-conversion was simply put directly into the interferometer instead.

Actually, Nagata et. al. spend considerable space discussing the type of loophole that we are addressing here. In their discussion, they show that if the probability $\eta$, to detect the four-photon state multiplied by the fringe visibility of this four-photon state does not exceed some value, then the SQL cannot be beaten. I already disagree with the basic analysis Nagata et. al. provide here. The theoretical detection probability of the state with four-fold phase sensitivity is less than $1/2$, therefore the SQL cannot be exceeded even assuming perfect fringe visibility and perfect detection efficiency. We have that four entangled photons provide at most a factor $1/\sqrt{4}$ improvement compared to the SQL, however for $\eta < 1/2$ the overall improvement is less than 1. It seems in their citation of ref (11), that Nagata et. al. again conflate measurement of observable $A$, with measurement of the quantum phase $\phi$ in performing this analysis. The minor issue of the theoretical impossibility of achieving any uncertainty improvement, Nagata et. al. then neglect to actually measure the probability to detect the four-photon state, and instead assume the theoretical value $\eta = 3/8$. This avoids the fact that the photo-detection efficiency of 0.6 greatly reduces the probability to detect the four-photon state. In particular, the probability to detect the four-photon state is $0.6 \times 0.6 \times 0.6 \times 0.6 = 0.13$. Compared to detection of the single photon state this is a probability reduction by 0.2. In addition, Nagata et. al. neglect to consider if the probability to generate the four-photon state is also reduced with respect to the probability to generate the single photon state.

\textbf{\textcolor{NavyBlue}{Analysis:}} The record efficiency for generation of photon pairs via spontaneous parametric downconversion is on the order of $10^{-6}$. If these $10^6$ photons which were used to generate photon pairs or quadruplets, were instead inserted into the interferometer, a much lower uncertainty would be achieved. Nagata et. al. report a generation rate of photon pairs of $1.7 \times 10^{-2}$ per pulse, and a generation rate of the four-photon state of $2.8 \times 10^{-4}$ per pulse, however the number of input photons per pulse is not given. Even if we exclude the poor efficiency per optical pulse and just look at the rate of single photon generation and detection compared to generation and detection of photon pairs or four-photon states, then a much lower uncertainty is achieved when employing single photons compared to the four-photon states. Fig. 3 shows that the single photon count rate is $1.5 \times 10^5$\,Hz compared to a four-photon count rate of $0.1$\,Hz. Therefore it is clear that even allowing for a factor $\sqrt{4}$ improvement (which Nagata et. al. do not demonstrate due to reduced visibility and non-optimal sensing state), the reduced probability to generate and detect the four-photon state does not outperform the uncertainty when all of the single photons are utilised. In this analysis we have not needed include the reduced visibility of the four-photon state, however the photo-detector efficiency is implicity included in photon count rate.

\noindent\textcolor{Maroon}{Uncertainty below $N = 1$ limit? The uncertainty that Nagata et. al. achieve with a four-photon entangled state is orders of magnitude worse than the uncertainty achieved with single photon input states in the same amount of time.}\\

\newpage

\subsection*{Interaction-based quantum metrology showing scaling beyond the Heisenberg limit. \emph{Nature} (2011) \href{https://doi.org/10.1038/nature09778}{10.1038/nature09778}}

Napolitano et. al. measure the magnetisation of an ensemble of $\sim10^6$ cold rubidium-87 atoms by passing detuned light through the ensemble and measuring the polarisation shift that the photons experience. Although no entanglement is exploited in their scheme, Napolitano et. al. claim to surpass both the SQL and the HL for estimating the atomic magnetisation as a function of the number of probe photons. This claim seems to be primarily based on a scaling argument, in that the uncertainty displays a better scaling with photon number in comparison to both the SQL and the HL. However, Napolitano et. al. claim multiple times during the manuscript to demonstrate an uncertainty that is below both the SQL and the HL, not just a superior scaling.\\

\textbf{Loophole \ref{item_scaling}:} An uncertainty that scales as $\Delta \hat{\phi} \propto \frac{1}{N^{3/2}}$ is demonstrated, but the absolute uncertainty does not go below $\Delta \hat{\phi} = \frac{1}{\sqrt{N}}$. As Napolitano et. al. measure a multiplicative factor in a time-independent Hamiltonian, their experimental uncertainty must be bounded by Eq.\,\ref{eq:SQL}, and for measurement of the atomic magnetisation $F_z$ in particular $\lambda = 1$, so we have (assuming an interaction time sufficient for absorption of each photon): $\Delta \hat{F}_z > 1/\sqrt{N}$.

\textbf{\textcolor{NavyBlue}{Analysis:}} Figure 3 shows the uncertainty that Napolitano et. al. demonstrate for estimating the atomic magnetisation. With $10^6$ photons the \emph{fractional} uncertainty in the atomic magnetisation is 10, giving an absolute uncertainty of $\Delta \hat{F}_z \approx 10^5$ or eight orders of magnitude worse than the SQL. Near the maximum photon number that Napolitano et. al. investigate with $5 \times 10^7$ photons, a \emph{fractional} uncertainty of $\sim 0.05$ is reported, giving an absolute uncertainty of $\Delta \hat{F}_z \approx 5 \times 10^3$, which is more than seven orders of magnitude worse than the SQL.

\noindent\textcolor{Maroon}{Uncertainty below $N = 1$ limit? No entanglement is exploited, so it is a moot point.}

\newpage








\subsection{Objections to the bound of Eq.\,\eqref{eq:qdyne}}\label{App:objections}
In achieving a frequency uncertainty below the T-E uncertainty limit with a single spin sensor, one might object that we are taking advantage of additional resources not available to other Heisenberg limited techniques. In particular, one might argue that the enhanced precision comes from measurement of a classical parameter rather than a quantum parameter, which allows the precision to be increased by performing more than one measurement. Another common objection is that the enhanced precision comes from access to a stable clock which is unavailable to other measurements. In addressing these objections, we need to precisely keep account of the resources that are utilized in defining the quantum uncertainty limits. This is a perilous task. First the basic resources necessary for any quantum measurement should be defined. As we are unaware of a complete definition, we attempt to provide a minimal list below. Further excess resources beyond this basic list that are generally required for quantum measurements that reach the standard quantum limit in estimating some classical parameter are accounted for below. Rather than perfectly quantify the physical amount of each resource, this following list is a first attempt at noting what resources are needed.

\subsubsection{Resources required for any quantum measurement}
\begin{enumerate}
\item A quantum system to be used as a sensor. Here we precisely count the number of atomic/indivisible components which the system contains and which are used to extract information about a physical parameter. We implicitly require that the state of the system can be controlled, readout and initialised. This first requirement demarcates our definition of a quantum measurement where -- a `quantum' system measures a `classical' parameter -- from another commonly used definition where -- a classical detector is used to measure a quantum system -- in quantum state tomography for example.
\item A physical parameter to be measured, which is assumed to have a fixed, constant and finite value in our treatment. Often this parameter considered to have some platonic existence which does not require physical resources, but this is an abstraction. A more precise treatment should additionally consider the number of resources that the measured parameter contains. For example, when measuring the amplitude of some magnetic field, a precise treatment of the uncertainty bounds should consider the number of photons in the magnetic field. The SQL (and HL) generally assumes that the field contains a large number of photons. Alternatively, one can also allow that the state of the quantum sensor or a particular eigenvalue of the sensor is the parameter to be measured. In this treatment, the state should be well defined (i.e. not a mixed state), and the bounds are not derived in terms of the number of particles, but rather the number of copies that are available. However, for simplicity we exclude this from our definition (see point 1 above).
\item An additional classical measuring device which is used to perform projective readout of the quantum sensor. Normally, these resources are not explicitly included and one assumes that perfect, instantaneous, projective measurements are possible in deriving the limits. However, physical resources are required to perform projective measurements, and in general, employment of more resources allows for faster measurements with higher fidelity.
\item A well defined spatial reference frame. This reference frame provides the measurement basis, see for example \cite{Bartlett2007}.
\item A clock. Just as a well defined spatial reference frame is required for a quantum measurement, we posit that a well defined measure of time is also required. Although not commonly discussed in the literature, we are unaware of any quantum measurement which does not implicitly use a stable clock. To me it is intuitively obvious, and should be taken as true that a clock is required, since it allows one to define a spatial reference frame at any given moment in time.\\
\end{enumerate}

\subsubsection{Additional resources required for quantum measurements in general and saturation of the SQL}
\begin{enumerate}
\item Control fields for controlling the evolution of the quantum sensor. These fields are additional to the fields which are required to readout and initialize the sensor. We assert that a well defined spatial frame and a clock are required to generate the control fields. As this possibility for control over the entire experimental duration is assumed in deriving quantum uncertainty limits, a stable clock is implicitly included as a requirement.
\item Perfect knowledge and control over all other parameters in the Hamiltonian.
\item No systematic errors. This relates to having perfect knowledge and control over all other elements of the Hamiltonian, but it can also be that systematic errors fundamentally arise even assuming perfect knowledge and control over all other parameters. In this case one would not be able to reach the SQL.
\item Spatial resources. The requirement for independent sensors in the SQL means that there is no interaction between sensors, therefore they need to be infinitely apart. Thus the classical parameter to be estimated should also have infinite spatial extent.
\item Resources to generate the parameter to be estimated, so that there is no absorption/destruction of the signal by the sensors, since increasing $N$ should not decrease signal seen by each atom. Can also be in the form of a low entropy source so that after each measurement of the parameter, the value of the parameter can be reinitialised (see Point 2).
\item Prior knowledge on the parameter to be estimated. For parameters that have an unbounded range of values, in general some prior information on the range of values that the parameter can take is required.
\item Computational resources to extract the best estimator from the measurement data-set.
\end{enumerate}

Even assuming all of these resources are available, it can be that no efficient estimator for the parameter exists which saturates the SQL.

\subsubsection{Comparison of resources needed for Ramsey interferometry and Qdyne type measurements in estimation of the oscillator frequency}
We compare estimation of the oscillator (e.g. a near-resonant microwave or laser field) frequency using Ramsey interferometry as performed with atomic clocks to estimation using a heterodyne measurement as implemented with the Qdyne technique. When comparing estimation strategies we endeavour to ensure that no extra resources are employed for one strategy that is not available to the other, and thus show that the same basic resources are available in each situation.\\

\textbf{Stable clock:}
We claim that with access to a single atom, and a (classical) clock that is less stable than the atomic frequency, one can outperform any state-of-the-art atomic clock in terms of absolute frequency uncertainty for some finite amount of time. The minimal requirement to achieve a lower frequency uncertainty than the atomic clock is that the classical clock has a minimum stability longer than $t_{\pi}$, thereby allowing more than one measurement to be performed. In addition to stability, the classical clock must also have a finite accuracy to allow generation of a control field at the correct frequency. The systematic error on the control frequency should be much better than $1/t_{\pi}$. In this case, if we have a total measurement time allowing one to perform several measurements, $T > t_{\pi}$ in Eq.\,\ref{eq:qdyne}, then an uncertainty below $1/T$ is obtained. The magnitude of reduction below the $1/T$ limit depends on the stability of the classical clock and the systematic error, but if the above requirements are satisfied, then we ensure that the T-E limit can be overcome.

We also assert that the same resources are required for atomic clocks. A stable clock, in addition to an atomic system with stable Larmor frequency is also needed for the duration of one Ramsey sequence in atomic clock operation. This clock is used to stabilise the microwave/laser source which acts as local oscillator to perform the $\pi/2$-pulses. Indeed the classical clock must be stable, not only for the time taken to do a single Ramsey sequence, but much longer since it takes several runs to accumulate enough statistics to estimate the local oscillator. This is because of quantum projection noise which cannot be overcome, meaning that the frequency of the local oscillator must be stable to much better than $1/t$ to allow efficient estimation. The clock used for this stabilisation is called the flywheel oscillator in atomic clocks, and its stability is often the limitation to $t$, and not the atomic dephasing time. Beyond stability, the clock must also have an accuracy or systematic error to much better than $1/t$ so that the correct fringe is identified. Finally, we have that $t$ must be much longer than $t_{\pi}$ in atomic clocks, in order to achieve approximately $1/T$ estimation uncertainty. We have thus shown that the clock used for Qdyne is not better than the clock required for atomic clock operation.

Although it is perhaps not clear, we are not claiming that with a less accurate clock, one can ever outperform atomic clocks in terms of fractional frequency uncertainty, which is the figure of merit for time standards. Rather this analysis addresses the absolute frequency uncertainty, which is less valuable for time standards, but perhaps a more appropriate figure of merit for spectroscopy. The resources required to achieve this uncertainty are exactly the same as the resources required by atomic clocks, they are simply put to more efficient use to achieve a slightly modified goal. We are also not claiming to achieve an absolute frequency uncertainty that outperforms atomic clocks at very long timescales. Due to systematic errors, the ultimate absolute uncertainty that is achievable in the limit of infinite measurement time will be better in atomic clocks. Note, the Cramer-Rao lower bound assumes that there are no systematic errors, and it turns out that systematic errors quickly become greater than statistical errors in many experiments. This is true for both atomic clocks and heterodyne type measurements. 

How about estimation of other physical parameters? Let's take estimation of the magnetic field amplitude using Ramsey spectroscopy. A stable clock is needed to set the phase of the final $\pi/2$-pulse with relation to the initial $\pi/2$-pulse. This timescale sets the maximum duration over which one can achieve the Heisenberg limit, in addition to the constraint set by a stable qubit frequency. Thus a stable clock is also needed for this strategy and the clock must be stable (and accurate) for at least time $T$ to achieve $1/T$ uncertainty reduction.\\

\textbf{Classical signal:}
A further objection could be that the `signal' that atomic clocks are measuring is really the atomic frequency and not some additional classical frequency. In this case, measurement back-action destroys the quantum state, and this prohibits one from achieving sub-linewidth estimation by performing more than one measurement. While that may be true, now we are comparing apples with oranges, since estimation of the Larmor frequency of a single atom is not equivalent to estimating the frequency of a classical field composed of many photons. This argument also neglects the fact that atomic clocks do indeed measure the frequency of a classical oscillator which is composed of many photons. I.e. the laser or microwave field which is resonant to the atomic transition and used as the distributed clock. Excluding this comparison and instead insisting we focus on the atomic Larmor actually introduces asymmetries that we are trying to avoid. In particular, it forces that we compare estimation of $\omega_0$ in one technique to estimation of $\omega$, so we are not even comparing estimation of the same physical parameters.

Put simply, atomic clocks estimate the frequency of some classical oscillating field using $N$ particles, the Qdyne technique also estimates the frequency of the same classical oscillating field using $N$ particles. It seems like a fair comparison to me.\\

\textbf{Prior information:}
The amount of prior information on the signal frequency required for Qdyne is much less than needed for atomic clocks, and much less than $1/T$. For Qdyne, the signal frequency should be known to well within $1/t_{\pi}$, which is a bigger prior uncertainty than $1/t \ll 1/t_{\pi}$ for atomic clocks. The amplitude of the signal is assumed to be known in Qdyne, but this is also assumed for atomic clocks, and both techniques have quadratic dependence on the amplitude of the signal field. The phase of signal does not need to be known for Qdyne just as in atomic clocks.\\




\textbf{Computational resources:}
One fair criticism could be that the computational resources required to achieve a reduced frequency uncertainty in Qdyne are much greater than required for atomic clocks. The reason for this is that the enhanced frequency uncertainty in Qdyne is obtained from analysis of a data-set containing many data-points. For $R \gg 1$ datapoints, an uncertainty improvement approximately $1/\sqrt{R}$ is obtained. A simple algorithm that achieves this uncertainty involves performing a fast-fourier transform, taking order $R \log(R)$ steps, performing a peak search, taking approximately $R/2$ steps, and then performing curve fitting on the peak to obtain sub-linewidth precision. If it ultimately turns out that the only advantage that entanglement provides for precision sensing is a reduction in computational resources to obtain an efficient estimator, this would in itself be a remarkable discovery.

\end{document}